\documentclass[journal]{IEEEtran}
\IEEEoverridecommandlockouts
\makeatletter
\def\endthebibliography{%
  \def\@noitemerr{\@latex@warning{Empty `thebibliography' environment}}%
  \endlist
}
\makeatother

\usepackage{cite}
\usepackage{amsmath,amssymb,amsfonts}
\usepackage{algorithm}
\usepackage{algpseudocode}
\usepackage{graphicx}
\usepackage{textcomp}
\usepackage[most]{tcolorbox}
\usepackage{xcolor}
\usepackage{balance}
\usepackage{multirow}
\usepackage{booktabs}
\usepackage{url}
\usepackage[caption=false]{subfig}
\usepackage[belowskip=-8pt,aboveskip=2pt]{caption}
\usepackage{float}
\usepackage{stfloats}
\usepackage{enumitem}
\usepackage[colorlinks,
            linkcolor=black,
            anchorcolor=black,
            citecolor=black]{hyperref}
\pdfoutput=1

\newcommand{\wb}{\textsc{WirelessBench}}
\newcommand{\wbr}{\textsc{WirelessBench-Ref}}

\ifodd 1

\newcommand{\com}[1]{{\color{blue}#1}}
\else

\newcommand{\com}[1]{}
\fi

\def\BibTeX{{\rm B\kern-.05em{\sc i\kern-.025em b}\kern-.08em
    T\kern-.1667em\lower.7ex\hbox{E}\kern-.125emX}}

\begin{document}

\title{WirelessBench: A Tolerance-Aware LLM Agent Benchmark for Wireless Network Intelligence}

\author{Jingwen Tong,~\IEEEmembership{Member,~IEEE},  Fang Liu, Linkai Xv, Shiliang Lu, Kangqi Li, Yiqian Zhang, Yijie Song, Zeyang Xue, and Jun Zhang,~\IEEEmembership{Fellow,~IEEE}
\thanks{Jingwen Tong, Fang Liu, Linkai Xv, Shiliang Lu, Kangqi Li, Yiqian Zhang, Yijie Song, and Zeyang Xue are with the College of Electronics and Information Engineering, Shenzhen University, Shenzhen, China (e-mails: eejwentong@szu.edu.cn; liuf@szu.edu.cn; \{2994034230, 2844686856, 372278887, yq-218, 3495753697,  2334929069\}@qq.com);
Jun Zhang is with the Department of Electronic and Computer Engineering, The Hong Kong University of Science and Technology, Hong Kong SAR (e-mail: eejzhang@ust.hk). (The corresponding author is Fang Liu)}
}

\maketitle

\begin{abstract}
LLM agents are emerging as a key enabler for autonomous wireless network management. Reliably deploying them, however, demands benchmarks that reflect real engineering risk. Existing wireless benchmarks evaluate single isolated capabilities and treat all errors uniformly, missing both cascaded-chain failures and catastrophic unit confusions (\textit{e.g.}, dB vs.\ dBm). We present \wb{}, the first tolerance-aware, tool-integrated benchmark for LLM-based wireless agents. \wb{} is organized as a three-tier cognitive hierarchy: domain knowledge reasoning (WCHW, 1{,}392 items), intent-driven resource allocation (WCNS, 1{,}000 items), and proactive multi-step decisions under mobility (WCMSA, 1{,}000 items). Moreover, \wb{} is established on three design principles: \emph{(i)}~tolerance-aware scoring with catastrophic-error detection; \emph{(ii)}~tool-necessary tasks requiring a 3GPP-compliant ray-tracing query for channel quality; and \emph{(iii)}~Chain-of-Thought (CoT)-traceable items, where every benchmark item ships with a complete CoT trajectory enabling fine-grained diagnosis of where in the reasoning chain an agent fails. Our numerical results show that the direct-prompting model (GPT-4o) scores $68\%$, trailing a tool-integrated agent ($84.64\%$) by $16.64$\,pp; $23\%$ of errors are catastrophic failures invisible to exact-match metrics. More importantly, the hierarchy decomposes errors into four actionable diagnostic categories that flat evaluation cannot reveal. Code and data: \url{https://wirelessbench.github.io/}.
\end{abstract}


\section{Introduction}
\label{sec:intro}

\IEEEPARstart{T}{he} growing complexity of next-generation wireless networks is driving a fundamental shift toward \emph{AI-native} network management~\cite{letaief2019roadmap}. As mobile systems evolve from 5G-Advanced to 6G, the operational decision space (i.e., spanning real-time spectrum coordination, dynamic network slicing, mobility-aware resource allocation, and proactive fault diagnosis) has expanded far beyond what human operators or classical optimization alone can manage~\cite{singh2025ai,tong2022age}. Large language model (LLM)-based AI agents, which combine natural-language understanding with external tool invocation, memory, and multi-step reasoning~\cite{wang2024survey_agent}, have emerged as a promising solution. They can interpret free-text network intents, invoke domain-specific tools, chain intermediate results, and produce structured operational outputs. If these agents prove reliable, they could automate the operational decision chains and accelerate both deployment speed and network intelligence. However, trusting an AI agent with live network decisions requires rigorous evaluation.

The research community has made substantial progress toward this vision along two complementary lines, establishing strong technical foundations for AI-driven wireless management. On the \emph{model side}, domain-adapted LLMs have demonstrated that considerable telecom knowledge can be internalized: TeleQnA~\cite{maatouk2024teleqna} showed that LLMs can recall 3GPP standards; WirelessLLM~\cite{shao2024wirelessllm} aligned LLM capabilities with wireless domain knowledge via prompting and retrieval; TelecomGPT~\cite{zou2024telecomgpt} and NetLLM~\cite{liu2024netllm} fine-tuned LLMs on telecom corpora; and rephrase-and-contrast training~\cite{rephrase2024contrast} with API-focused tuning~\cite{nefmind2025} further deepened domain reasoning. On the \emph{agent side}, the ReAct paradigm~\cite{yao2023react} enabled reasoning--acting loops, and a rapidly growing ecosystem of wireless AI agents has emerged: Agoran deploys cooperating AI agents in a 6G RAN marketplace~\cite{chatzistefanidis2025agoran}; AgentRAN constructs self-organizing agent hierarchies for Open~RAN~\cite{elkael2025agentran}; agentic TinyML places lightweight agents at network edges~\cite{saleh2025agentic_tinyml}; SignalLLM creates general-purpose agents for signal processing~\cite{ke2025signalllm}; and further systems~\cite{lin2025llm_network,comagent2025,mxai2025,scnoc_agentic_2025} broaden the scope from intent-driven automation to satellite agentic control. Together, these advances demonstrate that building capable wireless AI agents is increasingly \emph{feasible} and \emph{urgent}.

However, amid rapid agent development, a \emph{critical gap} has been overlooked: how to \emph{evaluate} agents in a manner that reflects the actual risk profile of wireless deployment. A growing body of wireless/telecom benchmarks has been recently developed: TeleQnA~\cite{maatouk2024teleqna} and GSMA Open-Telco~\cite{gsma2025_otellm} assess knowledge recall through QA tasks; TeleMath~\cite{telemath2025}, WirelessMathBench~\cite{wirelessmathbench2025}, and WirelessMathLM~\cite{wirelessmathlm2025} test mathematical reasoning in isolation; 6G-Bench~\cite{yang2026_6gbench} adds semantic-level and network-level reasoning. However, these benchmarks share two structural limitations that render them insufficient for deployment-readiness assessment. \emph{First}, they evaluate capabilities on \emph{single isolated axes} (i.e., knowledge recall \emph{or} formula computation). Real deployment failures typically arise from cascaded-chain failures, and evaluating each link in isolation overlooks the most deployment-relevant risk. \emph{Second}, existing metrics \emph{conflate benign approximation with catastrophic error}. For instance, a single unit confusion between dBW and dBm can cause a $1{,}000\times$ power misestimate and trigger coverage outage~\cite{truong2025fantastic}. Therefore, it is critical to build a tolerance-aware LLM agent benchmark for wireless networks.

Our ongoing research program on LLM-based wireless agents underscores this evaluation gap. In WirelessAgent~\cite{tong2024wirelessagent}, we introduced a ReAct-style agent framework equipped with domain-specific tools (\textit{e.g.}, a telecom formula retriever, a precision calculator, and a ray-tracing channel predictor) that achieved near-optimal performance on network slicing tasks. In WirelessAgent$^{++}$~\cite{tong2025wirelessagentpp}, we moved beyond manual workflow design by casting agent construction as a program search problem solved through domain-adapted Monte Carlo Tree Search (MCTS), automatically discovering task-adaptive workflows that outperform hand-crafted baselines by up to $31\%$ and general-purpose optimizers by $11.9\%$. Through this agent-building experience, we found that \emph{evaluation was the principal bottleneck}: existing benchmarks provided only flat aggregate scores and could not diagnose \emph{whether} an agent failed due to knowledge gaps, tool-use errors, reasoning-chain breaks, or catastrophic unit mistakes. Without such diagnostic resolution, improving agent reliability remained ad hoc rather than systematic.

In this work, we release \wb{}, a tolerance-aware and tool-integrated benchmark for AI agents in wireless networks.
\wb{} comprises three complementary tasks: Wireless Communication Homework (WCHW, 1,392 samples) for domain knowledge reasoning, Wireless Communication Network Slicing (WCNS, 1,000 samples) for intent-driven resource allocation, and Wireless Communication Mobile Service Assurance (WCMSA, 1,000 samples) for proactive multi-step decisions under mobility.
\begin{itemize}[leftmargin=*]
  \item \textbf{Knowledge Reasoning (WCHW):} 1,392 problems testing application of wireless fundamentals (Shannon capacity, BER, modulation, coding, fading channel analysis) from classical communication textbooks and research papers.
  \item \textbf{Intent-to-Allocation (WCNS):} 1,000 problems where the agent classifies a free-text service request, \emph{calls a ray-tracing tool} to obtain the channel quality indicator (CQI) at the user's location on a campus-scale radio map~\cite{wen2025neural}, and produces a four-field structured allocation (slice type, CQI, bandwidth, throughput).
  \item \textbf{Proactive Decision Chain (WCMSA):} 1,000 problems where the agent predicts a mobile user's future position via trajectory analysis, \emph{invokes ray tracing at the predicted location}, and produces a six-field decision (position, CQI, slice type, bandwidth, throughput, QoS feasibility).
\end{itemize}

Building a benchmark that faithfully reflects operational risk, however, is far from trivial. We follow three design principles to guide \wb{}. \emph{First}, unlike single-skill QA benchmarks, \wb{} introduces \textbf{tolerance-aware scoring with catastrophic-error detection}. Numeric outputs are evaluated under tiered relative-error thresholds, while unit mismatches (e.g., dB vs.\ dBm, kbps vs.\ Mbps) and order-of-magnitude errors are explicitly flagged and penalized to zero. This makes the metric operationally grounded. \emph{Second}, WCNS and WCMSA tasks are \textbf{tool-necessary design} on a campus-based radio map conforming to 3GPP TR~38.901. This turns \wb{} from a QA test into a genuine \emph{agentic} evaluation, where the agent must autonomously decide when to invoke tools and integrate tool outputs into downstream reasoning. \emph{Third}, every benchmark item is accompanied by a \textbf{complete CoT trajectory} that characterizes the expected reasoning path at each tier, enabling fine-grained diagnosis of whether a failure stems from formula misuse, a missing tool call, or cross-branch error propagation.

The resulting evaluation yields risk-relevant diagnostic information that conventional benchmarks cannot provide. Across the full 3,392-sample release, our experiments reveal three key findings. (i)~\emph{Tool access is a dominant capability factor}: the best direct-prompting frontier model (GPT-4o) achieves only $68\%$ on average, while a tool-integrated reference pipeline reaches $84.64\%$, a $16.64$\,pp gap that quantifies the performance ceiling imposed by the absence of tool orchestration. We note that this gap partly reflects information asymmetry (the reference pipeline has ray-tracing access that direct prompting lacks; see Section~\ref{sec:discussion} for analysis). (ii)~\emph{Tolerance-aware scoring exposes hidden risk}: tolerance-aware metrics recover $16.18$\,pp of benign approximation for direct prompting but only $5.59$\,pp for tool-verified agents, revealing that $23\%$ of all errors are catastrophic unit/magnitude failures \emph{invisible} to exact-match metrics. (iii)~\emph{Cascaded chains expose failure propagation}: the three-tier hierarchy decomposes errors into formula misapplication ($31\%$), reasoning-path breaks ($28\%$), unit/magnitude confusion ($23\%$), and arithmetic errors ($18\%$).

The main contributions of this work are summarized as follows:
\begin{itemize}[leftmargin=*]
  \item We present \wb{}, a tolerance-aware and tool-integrated benchmark for evaluating the \emph{reliability} of LLM-based AI agents in mobile-network management tasks. To our knowledge, it is the first wireless benchmark to jointly integrate tolerance-aware scoring, tool-necessary tasks, and CoT-traceable items, covering the operational decision chain from domain knowledge through tool-augmented allocation to proactive QoS verification.
  \item We introduce tolerance-aware scoring with explicit catastrophic-error detection, providing risk-sensitive metrics that separate acceptable approximation from operationally dangerous mistakes, and enabling mobile-network operators to identify agents that may cause coverage outages or resource misallocation.
  \item We design tool-necessary evaluation tasks with pre-computed 3GPP-compliant ray-tracing propagation data, establishing a wireless benchmark that inherently requires agentic tool use for CQI acquisition and position-dependent channel estimation.
  \item We develop a data-construction pipeline combining multi-model grading, statistical item filtering, AI-assisted diagnosis, and human verification, raising benchmark quality and auditability beyond one-pass data generation.
\end{itemize}

The remainder of this paper is organized as follows. Section~\ref{sec:related} reviews related work. Section~\ref{sec:design} presents the design principles and architecture of \wb{}. Section~\ref{sec:pipeline} details the data construction and quality-assurance pipeline. Section~\ref{sec:calibration} introduces difficulty calibration and reliability analysis. Section~\ref{sec:experiments} reports evaluation protocols and empirical findings. Section~\ref{sec:discussion} discusses implications and limitations, and Section~\ref{sec:conclusion} concludes the paper.

\section{Related Work}
\label{sec:related}

\begin{table*}[t]
\centering
\caption{Comparison of wireless/telecom AI evaluation benchmarks. \textbf{Tol.}\,=\,tolerance-aware scoring; \textbf{Cat.}\,=\,catastrophic error detection; \textbf{Tool}\,=\,tool-necessary tasks; \textbf{Chain}\,=\,multi-step decision chain; \textbf{CoT}\,=\,CoT-traceable trajectory. \checkmark\,/\,-- indicates supported/not supported.}
\label{tab:bench_compare}
\resizebox{\textwidth}{!}{
\begin{tabular}{lccccccccc}
\toprule
\textbf{Benchmark} & \textbf{Scope} & \textbf{Size} & \textbf{Answer Type} & \textbf{Tol.} & \textbf{Cat.} & \textbf{Tool} & \textbf{Chain} & \textbf{CoT} &  \\
\midrule
TeleQnA~\cite{maatouk2024teleqna} & Knowledge QA & 10K+ & MCQ / Short text & -- & -- & -- & --  & -- \\
GSMA Open-Telco~\cite{gsma2025_otellm} & Knowledge QA & -- & MCQ / Free-form & -- & -- & --  & -- & -- \\
TeleMath~\cite{telemath2025} & Math reasoning & 500 & Numeric & -- & -- & -- & --  & -- \\
WirelessMathBench~\cite{wirelessmathbench2025} & Math modeling & 2.8K & Numeric & -- & -- & --  & -- & Partial \\
WirelessMathLM~\cite{wirelessmathlm2025} & Math reasoning (RL) & 4K & Numeric & -- & -- & -- & --  & Partial \\
6G-Bench~\cite{yang2026_6gbench} & Semantic + Network & 3,722 & Text / Structured & -- & -- & -- & Partial  & -- \\
Go Gentle~\cite{go2025_finaldance} & Telecom eval. & -- & Mixed & -- & -- & -- & -- & -- \\
\midrule
\textbf{WirelessBench (Ours)} & \textbf{3-tier (K+I+D)} & \textbf{3,392} & \textbf{Num / Formula / Struct} & \checkmark & \checkmark & \checkmark & \checkmark & \checkmark \\
\bottomrule
\end{tabular}}
\end{table*}

\subsection{LLMs and AI Agents for Wireless Networks}
LLM applications in wireless networks have evolved through three successive phases. The first phase focused on \emph{knowledge injection}: TeleQnA~\cite{maatouk2024teleqna} evaluated how well LLMs recall telecom standards; WirelessLLM~\cite{shao2024wirelessllm} aligned LLMs with wireless domain knowledge through prompting and retrieval; NetLLM~\cite{liu2024netllm} fine-tuned LLMs for networking tasks; and TelecomGPT~\cite{zou2024telecomgpt}, along with rephrase-and-contrast training~\cite{rephrase2024contrast} and API-focused fine-tuning~\cite{nefmind2025}, deepened domain reasoning. These contributions demonstrate that LLMs can internalize substantial wireless knowledge, but they do not evaluate the agent's ability to \emph{use that knowledge} in multi-step workflows.

The second phase introduced \emph{agentic reasoning and tool use}. Building on the ReAct architecture~\cite{yao2023react} and the broader agent landscape~\cite{xi2023rise,wang2024survey_agent}, wireless AI agents have evolved from the AI-router concept~\cite{jiang2020intelligent_agent} to a rapidly expanding ecosystem spanning RAN marketplace coordination~\cite{chatzistefanidis2025agoran}, agentic observability~\cite{mxai2025}, self-organizing hierarchies~\cite{elkael2025agentran}, edge-deployed TinyML agents~\cite{saleh2025agentic_tinyml}, signal-processing agents~\cite{ke2025signalllm}, and diverse network-control, RAG-based, and intent-driven systems~\cite{lin2025llm_network,comagent2025,rag_rrm_2025,scnoc_agentic_2025,intent_based_6g_2025,smallmodels2025,icwlm_2025}. In addition, our WirelessAgent~\cite{tong2024wirelessagent} and WirelessAgent$^{++}$~\cite{tong2025wirelessagentpp} works contributed tool-integrated frameworks for wireless networks, yet both confirmed that \emph{evaluation} was the real bottleneck.

This rapid system-side progress creates an urgent demand for the \emph{third and currently missing phase}: evaluation that is \emph{tolerance-aware} (separating benign approximation from catastrophic errors), \emph{tool-integrated} (testing agentic interaction with external engines), and \emph{CoT-traceable} (every benchmark item ships with a complete CoT trajectory, enabling fine-grained diagnosis of where an agent fails). \wb{} is designed to fill these evaluation gaps.

\subsection{Wireless/Telecom Benchmarking: Progress and Gaps}
Table~\ref{tab:bench_compare} summarizes the landscape. TeleQnA~\cite{maatouk2024teleqna} and GSMA Open-Telco~\cite{gsma2025_otellm} assess \emph{knowledge recall} via QA tasks but test neither mathematical precision nor multi-step reasoning, and do not distinguish benign approximation from catastrophic error. TeleMath~\cite{telemath2025}, WirelessMathBench~\cite{wirelessmathbench2025}, and WirelessMathLM~\cite{wirelessmathlm2025} target \emph{formula-level reasoning} but evaluate each computation in isolation with binary scoring (i.e., no tool interaction and tolerance-aware grading). 6G-Bench~\cite{yang2026_6gbench} broadens scope to semantic- and network-level reasoning; Go Gentle~\cite{go2025_finaldance} emphasizes practical telecom tasks. Neither, however, addresses tool integration or error-type differentiation.

Despite this progress, existing benchmarks share four common limitations: \emph{(i) None requires the model to interact with external tools} (e.g., ray-tracing propagation engines, CQI lookup services, or mobility predictors) as part of the evaluation. \emph{(ii) None implements cascaded multi-step decision chains} in which an error in one field (e.g., CQI) propagates to downstream fields (bandwidth, throughput, QoS), so propagation risk is invisible. \emph{(iii) None employs tolerance-aware metrics that explicitly distinguish benign approximation from catastrophic engineering errors} such as unit confusion or order-of-magnitude mismatches. \emph{(iv) None provides complete CoT trajectories} for benchmark items, preventing fine-grained diagnosis of where in the reasoning chain an agent fails. \wb{} targets all four gaps through unified multi-tier tasks, tool-necessary evaluation, tolerance-aware structured scoring, and CoT-traceable item design. We note that tolerance-aware scoring and multi-step evaluation are established ideas in general-purpose AI benchmarks (e.g., tiered scoring in math benchmarks, tool-use evaluation in agent benchmarks); the contribution here is their principled domain adaptation to the specific error taxonomy and operational risk profile of wireless engineering.

\subsection{Benchmark Reliability and Quality Assurance}
Benchmark reliability has become a first-order concern in AI evaluation: Truong~et~al.~\cite{truong2025fantastic} demonstrated that even widely used benchmarks contain annotation errors, ambiguous items, and near-duplicate questions that can distort model rankings and weaken scientific conclusions. While psychometric principles (e.g., item discriminability, scalability, and internal consistency) are standard in educational assessment, they remain largely absent from LLM benchmark construction, particularly in domain-specific settings.

This gap is especially damaging in wireless evaluation. The numeric, formula-heavy, and unit-sensitive nature of the domain means that a miskeyed unit can systematically favor models. \wb{} therefore incorporates a multi-stage psychometric-inspired pipeline, shifting construction from one-pass generation to a measurement-informed process. We emphasize that the model panel is far smaller than the hundreds of examinees typical in classical psychometrics; we use the term \emph{psychometric-inspired} to signal that the statistical indicators serve as quality heuristics rather than providing the same guarantees as full-scale psychometric calibration.

\section{WirelessBench: Design Principles and Architecture}
\label{sec:design}

\begin{figure*}[t]
\centering
\includegraphics[width=0.85\linewidth]{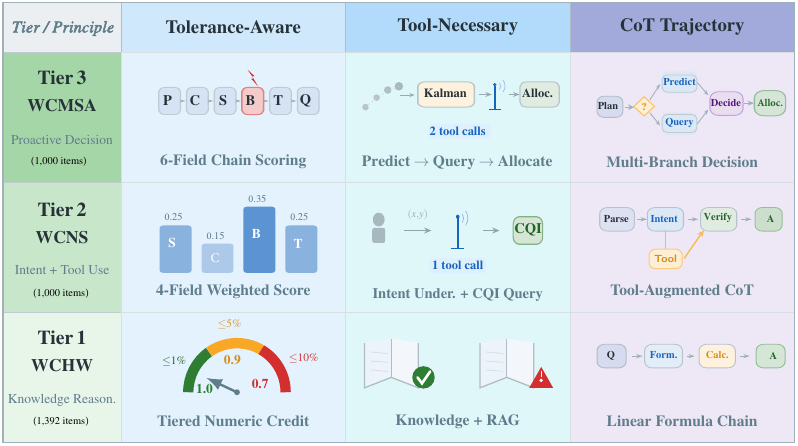}
\caption{\wb{} design framework organized as a $3\!\times\!3$ matrix. The vertical axis represents a three-tier cognitive hierarchy: Tier~1 WCHW (knowledge reasoning, 1{,}392 items), Tier~2 WCNS (intent understanding and tool use, 1{,}000 items), and Tier~3 WCMSA (proactive multi-step decision, 1{,}000 items). This framework/matrix increases complexity from bottom to top. The horizontal axis captures three cross-cutting design principles: \emph{Tolerance-Aware} scoring (tiered numeric credit, weighted field scores, and chain scoring), \emph{Tool-Necessary} evaluation (knowledge retrieval, CQI query, and multi-tool orchestration), and \emph{CoT Trajectory} analysis (linear formula chains, tool-augmented reasoning, and multi-branch decision paths). Each cell instantiates how a given principle manifests at a specific tier.}
\label{fig:design_framework}
\end{figure*}

Fig.~\ref{fig:design_framework} presents an overview of the \wb{}, organized as a $3\!\times\!3$ matrix that jointly captures \emph{what} is evaluated and \emph{how} it is evaluated.
The vertical axis represents a three-tier cognitive hierarchy (i.e., from Tier~1 knowledge reasoning (WCHW) through Tier~2 intent-driven tool use (WCNS) to Tier~3 proactive multi-step decision-making (WCMSA)), reflecting the progression from factual recall to chained operational decisions under mobility.
The horizontal axis introduces three cross-cutting design principles:
(i)~\emph{Tolerance-Aware} scoring, which replaces brittle exact-match metrics with tiered numeric credit and weighted composite scores;
(ii)~\emph{Tool-Necessary} evaluation, which mandates interaction with a physically grounded ray-tracing digital twin, escalating from knowledge retrieval through single-tool CQI query to multi-tool orchestration; and
(iii)~\emph{CoT Trajectory} analysis, which characterizes the expected reasoning path at each tier: linear formula chains for Tier~1, tool-augmented chains for Tier~2, and multi-branch decision trajectories for Tier~3.
The following subsections elaborate on each axis and cell of this matrix.

\subsection{Design Philosophy: Three-Tier Cognitive Hierarchy}
WirelessBench is built on a three-tier cognitive hierarchy that mirrors how AI systems are expected to operate in mobile networks. This design is motivated by a key observation: real wireless deployment failures rarely stem from a single missing fact; they arise when an agent cannot chain steps correctly, i.e., from domain knowledge through intent interpretation to operational decision.
\begin{itemize}[leftmargin=*]
  \item \textbf{Tier~1 (Knowledge Reasoning):} It verifies whether models can correctly apply wireless fundamentals such as Shannon capacity, BER computation, SNR conversion, modulation/coding, and fading analysis. This tier provides a \emph{knowledge floor}: an agent that fails here cannot be trusted with downstream operational tasks.
  \item \textbf{Tier~2 (Intent Understanding \& Allocation):} It verifies whether models can map free-text service semantics to network intent, query external tools for channel information, and produce valid multi-field resource-allocation decisions. This tier tests the coupling of language understanding with tool use and domain computation.
  \item \textbf{Tier~3 (Proactive Multi-Step Decision):} It verifies whether models can perform chained, proactive reasoning under mobility, including trajectory-based position prediction, tool-based channel estimation at the predicted location, slice/bandwidth allocation, and QoS feasibility verification. This tier represents the most deployment-relevant capability: predicting network conditions and acting preemptively.
\end{itemize}
The hierarchy avoids the ``single-score illusion'' of flat QA benchmarks and enables \emph{capability decomposition}: a model may excel at Tier~1 knowledge but fail at Tier~3 chained decisions, a distinction that flat aggregate scores obscure. Importantly, the architecture is \emph{modular by design}: each tier is an independent task module with its own data format, scoring rubric, and tool requirements. This modularity means that additional task modules, such as multi-agent coordination, mmWave link management, or non-terrestrial-network handover, can be incorporated without restructuring the evaluation protocol, making \wb{} an extensible suite that grows with the community's needs.

\subsection{Task Taxonomy}
\textbf{WCHW (Wireless Communication Homework).} WCHW targets Tier~1 knowledge reasoning. The 1,392 problems span nine knowledge categories: modulation/demodulation, digital communication, analog communication, information theory, wireless channels, noise analysis, multiplexing, multiple access and cellular systems, and error-control coding. Expected outputs include numeric values with units, mathematical formulas, scientific-notation quantities, and short technical text. Multiple answer types within a single benchmark ensure that evaluators exercise numeric tolerance, formula normalization, and text-matching capabilities simultaneously.

\textbf{WCNS (Wireless Communication Network Slicing).} WCNS targets Tier~2 intent understanding and allocation with tool use. Each of the 1,000 samples provides a network state (slice capacities, active user counts), a user position in local coordinates, and a free-text service request (e.g., ``I need power-grid fault detection'' $\rightarrow$ URLLC). Critically, the user's CQI is \emph{not supplied in the prompt}; the agent must call a \texttt{ray\_tracing} tool with the user's $(x,y)$ coordinates and campus region to obtain CQI, then produce four structured outputs: slice type, CQI, bandwidth (via proportional fairness), and throughput (via the 3GPP TS 38.214 spectral-efficiency table). This task is \emph{tool-necessary}: an agent that skips the ray-tracing call or fabricates a CQI value cannot produce a correct slice allocation and throughput calculation.

\textbf{WCMSA (Wireless Communication Mobile Service Assurance).} WCMSA targets Tier~3 proactive multi-step decisions. Each of the 1,000 samples provides a historical trajectory (typically 4--6 positions), base-station parameters, network state, a service request with a minimum-rate QoS requirement, and latency sensitivity. The agent must: (1) predict the user's future position from the trajectory (e.g., via Kalman filtering), (2) call \texttt{ray\_tracing} at the \emph{predicted} position to obtain the future CQI, (3) classify the service intent, (4) allocate bandwidth, (5) compute throughput, and (6) verify QoS feasibility. The six-field structured output and the two tool dependencies (position prediction + ray tracing) make WCMSA the most challenging tier: errors compound through a chain of six interdependent fields.

Therefore, the three tasks provide progressive evaluation from formula-level reasoning to operational, mobility-aware decision chains, and the modular task-module design allows the suite to expand as new deployment scenarios emerge.

\subsection{Tolerance-Aware Scoring Framework}
Conventional exact-match metrics are insufficient for wireless engineering tasks because they conflate two fundamentally different failure modes: \emph{benign approximation} (minor rounding) and \emph{catastrophic engineering error} (unit confusion, order-of-magnitude mismatch). A 0.5\% throughput deviation is operationally irrelevant; confusing dBm with dBW yields a 30\,dB ($1,000\times$) power error that can cause coverage outage. \wb{} uses a tolerance-aware scoring framework to explicitly separate these cases.

\textbf{Numeric tolerance tiers.} For scalar outputs, let $\hat{y}$ be the prediction and $y$ be the reference. The relative error is
\begin{equation}
\epsilon = \frac{|\hat{y}-y|}{\max(|y|,\delta)},
\end{equation}
where $\delta$ is a small constant preventing division by zero. We use tiered credit:
\begin{equation}
s_{\mathrm{num}}(\epsilon)=
\begin{cases}
1.0, & \epsilon \le 0.01,\\
0.9, & 0.01 < \epsilon \le 0.05,\\
0.7, & 0.05 < \epsilon \le 0.10,\\
0.0, & \epsilon > 0.10.
\end{cases}
\end{equation}

\textbf{Catastrophic error detection.} Unit inconsistency and order-of-magnitude mismatches are explicitly detected and penalized. For instance, dBW/dBm misuse receives zero credit even when token-level similarity is high.

\textbf{Formula/text partial credit.} Formula answers are normalized (symbol and format) and scored using structure-aware similarity. Text answers are graded by weighted keyword and semantic consistency instead of strict lexical equality.

\textbf{Composite structured-task scoring.} For WCNS and WCMSA, field-level scores are aggregated by weighted sums to reflect the operational impact of each field.

Next, we introduce the detailed evaluation metrics for the WCHW, WCNS, and WCMSA tasks.

\textbf{WCHW.} A multi-type answer classifier first identifies the output type (numeric/scientific, formula, or text) and routes each answer to a dedicated evaluator with type-appropriate scoring. The task-level score is the mean over all samples.

\textbf{WCNS.} A four-field weighted composite score is used:
\begin{equation}
S_{\text{WCNS}}=w_1 S_{\text{slice}}+w_2 S_{\text{cqi}}+w_3 S_{\text{bw}}+w_4 S_{\text{tp}},
\end{equation}
with default weights $w_1=0.25,\,w_2=0.15,\,w_3=0.35,\,w_4=0.25$, prioritizing allocation quality (bandwidth) and throughput consistency. These weights are informed by engineering judgment (\textit{i.e.,} bandwidth allocation has the largest downstream impact on throughput), rather than derived from a formal standard; we discuss their sensitivity in Section~\ref{sec:discussion}.

\textbf{WCMSA.} A six-field weighted composite score is used:
\begin{equation}
S_{\text{WCMSA}}=\sum_{k=1}^{6} v_k S_k,
\end{equation}
with default weights $v = (0.15, 0.15, 0.20, 0.25, 0.20, 0.05)$ for position, CQI, slice type, bandwidth, throughput, and QoS verification, respectively. Position accuracy uses continuous distance-based decay:
\begin{equation}
S_{\text{pos}}=\max\left(0,\,1-\left(\frac{d(\hat{\mathbf{p}},\mathbf{p})}{d_{\max}}\right)^{1.2}\right),
\end{equation}
with $d(\cdot)$ the Euclidean distance and $d_{\max}=20$\,m (chosen so that prediction errors exceeding a typical small-cell radius receive zero credit), providing a gradient signal that distinguishes small improvements in mobility prediction. The exponent $1.2$ yields slightly super-linear penalty and was selected empirically; a sensitivity study of both $d_{\max}$ and the exponent is noted as future work in Section~\ref{sec:discussion}. Fig.~\ref{fig:scoring_weights} visualizes the field-level scoring weight distributions for all three tasks.

\subsection{Tool-Necessary Tasks}
\label{sec:digital_twin}
For the WCNS and WCMSA tasks, we construct the propagation environment from OpenStreetMap (OSM) data of the HKUST campus, covering three regions (North, Center, South). For each region, we pre-compute ray-tracing propagation results on a dense spatial grid, recording received power ($P_{\text{rx}}$, dBm), path loss (dB), signal-to-noise ratio (dB), line-of-sight (LOS) status, and the 3GPP CQI index derived from the SNR-to-CQI mapping specified in TS~38.214 Table 5.2.2.1-3.

The evaluated agent accesses this data through a \texttt{ray\_tracing(x, y, region)} tool call, which returns the CQI at the queried location. For WCNS, the agent receives the user's actual position and must invoke the tool once. For WCMSA, the agent must \emph{first predict} the user's future position and then invoke the tool at the predicted coordinates; an incorrect position prediction therefore cascades to an incorrect CQI and downstream allocation.

This architecture serves two purposes. \emph{First}, it enforces a minimal agentic capability: the model must recognize that CQI is missing, decide to call the tool, formulate correct arguments, and integrate the returned value. This separates \wb{} from QA benchmarks, where all information is self-contained in the prompt. \emph{Second}, the propagation data is physically grounded. It reflects actual campus geometry, building obstructions, and 3GPP-compliant path-loss models (TR~38.901 Urban Micro). This ensures that label values are operationally meaningful rather than randomly assigned. Full ray-tracing parameters and the 3GPP CQI-to-spectral-efficiency mapping table are documented in Appendices~\ref{app:digital_twin} and~\ref{app:cqi}, respectively.

\subsection{CoT Trajectory Design}
\label{sec:cot_trajectory}
Beyond scoring and tool usage, \wb{} explicitly characterizes the \emph{CoT trajectory} expected at each tier, providing a principled lens to analyze not only whether an agent reaches the correct answer but \emph{how} it reasons toward it (see the rightmost column of Fig.~\ref{fig:design_framework}).

\textbf{Tier~1: Linear Formula Chain.}
WCHW problems admit a simple sequential reasoning path: the agent reads the question~(Q), selects the appropriate formula (Form.), executes the calculation~(Calc.), and produces the answer~(A). The trajectory $\text{Q}\!\rightarrow\!\text{Form.}\!\rightarrow\!\text{Calc.}\!\rightarrow\!\text{A}$ is deterministic and unbranched; errors typically originate from formula misselection or unit confusion rather than from control-flow complexity.

\textbf{Tier~2: Tool-Augmented CoT.}
WCNS introduces a \emph{tool detour} into the reasoning chain: the agent first parses the service request~(Parse), identifies the network intent~(Intent), recognizes that CQI is missing and invokes the ray-tracing tool~(Tool), verifies consistency between the tool output and the task context~(Verify), and finally produces the four-field allocation answer~(A). The trajectory $\text{Parse}\!\rightarrow\!\text{Intent}\!\rightarrow\!\text{Tool}\!\rightarrow\!\text{Verify}\!\rightarrow\!\text{A}$ is still largely linear but contains a critical branch point: the decision to call the tool. Agents that skip this branch---hallucinating a CQI value instead---follow a shorter but incorrect trajectory and incur systematic throughput errors.

\textbf{Tier~3: Multi-Branch Decision.}
WCMSA requires a genuinely branching trajectory. Starting from a plan~(Plan), the agent faces a decision point (denoted ``?'' in Fig.~\ref{fig:design_framework}): it must \emph{simultaneously} predict the future position from the mobility trace~(Predict) and query the digital twin at the predicted location~(Query), then merge both branches to decide the resource allocation~(Decide) and output the six-field result~(Alloc.). The trajectory $\text{Plan}\!\rightarrow\!?\!\rightarrow\!\{\text{Predict},\text{Query}\}\!\rightarrow\!\text{Decide}\!\rightarrow\!\text{Alloc.}$ introduces \emph{error propagation across branches}: an inaccurate position prediction cascades through the query branch, producing a wrong CQI that corrupts the downstream allocation and QoS verification. This multi-branch structure is the key differentiator of Tier~3 and explains why aggregate scores drop sharply relative to Tier~2.

By making the expected CoT trajectory explicit, \wb{} enables fine-grained failure diagnosis: evaluators can pinpoint whether an error stems from formula misuse (Tier~1), a missing tool call (Tier~2), or cross-branch error propagation (Tier~3), moving beyond aggregate accuracy toward actionable capability profiling.

\section{Data Construction Pipeline}
\label{sec:pipeline}
\begin{figure*}[t]
    \centering
    \includegraphics[width=1\linewidth]{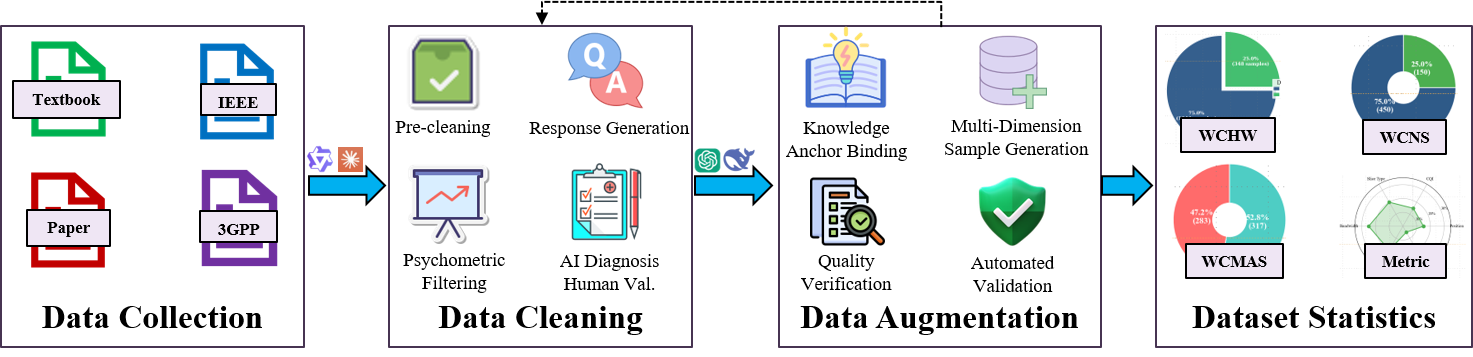}
    \caption{WirelessBench construction pipeline, including {Data Collection}, {Data Cleaning}, {Data Augmentation}, and {Dataset Statistics} modules.}
    \label{fig:overview}
\end{figure*}

Fig.~\ref{fig:overview} illustrates our four-stage pipeline to construct \wb{}. For each task, we begin with \textbf{Data Collection} from authoritative sources, including standard textbooks on wireless communications, research papers, and 3GPP standards. The \textbf{Data Cleaning} stage employs multi-model consensus: GPT-4, Claude, Qwen, and DeepSeek independently verify solutions, with disagreements flagged for expert review. We apply psychometric filtering (item-total correlation, Mokken scale analysis) to remove low-quality items. \textbf{Data Augmentation} uses human-AI collaboration where domain experts guide LLMs to generate variations while preserving semantic validity. The final \textbf{Dataset Statistics} show balanced coverage across tasks with comprehensive evaluation metrics.

\subsection{Seed Data Collection}
\textbf{WCHW seeds.} We synthesize and curate textbook-inspired wireless exercises across nine knowledge domains, including modulation/demodulation (AM, FM, PM, QAM, PSK, FSK), information theory (Shannon capacity, channel coding bounds), digital communication (BER, Nyquist rate, PCM/quantization), analog communication (Carson's rule, noise figure), wireless channels (Rayleigh/Rician fading, path loss), and system design (matched filtering, multiplexing, multiple access). Each seed item stores the question, a canonical numeric/formula/text answer, and a CoT solution trace with explicit intermediate steps to enable internal verification. We note that the CoT trajectory is critical for AI agents to improve their reasoning capabilities during the training phase.

\textbf{WCNS/WCMSA seeds.} We construct scenario-driven items grounded in two pillars of physical fidelity. \emph{First}, all CQI-to-spectral-efficiency mappings are anchored to 3GPP TS~38.214 Table 5.2.2.1-3 (CQI 1--15, QPSK through 64QAM, spectral efficiencies 0.15--5.55\,bps/Hz). \emph{Second}, channel conditions are derived from a campus-scale radio map of the HKUST campus: we import OpenStreetMap building geometry for three regions (North, Center, South), and pre-compute ray-tracing propagation at 3.5\,GHz carrier frequency using the 3GPP TR~38.901 Urban-Micro model with 30\,kHz subcarrier spacing. The resulting label set (received power, path loss, SNR, LOS status, and CQI) is stored in per-region CSV tables and served through a \texttt{ray\_tracing} tool interface. For WCMSA, mobility traces are generated following 3GPP specifications with pedestrian (1--2\,m/s) and vehicular (5--15\,m/s) mobility patterns, and position prediction labels are computed using a Kalman filter with state vector $[x, y, v_x, v_y]^\top$, process noise $\sigma_q = 0.5$\,m, and measurement noise $\sigma_r = 0.1$\,m. This end-to-end grounding ensures that generated labels are \emph{physically plausible and operationally interpretable}, not merely randomly assigned values. The task-specific data collection prompts are provided in Appendix~\ref{app:prompts}.

\subsection{Psychometric Data Cleaning Framework}
\begin{figure*}[t]
\centering
\includegraphics[width=0.90\linewidth]{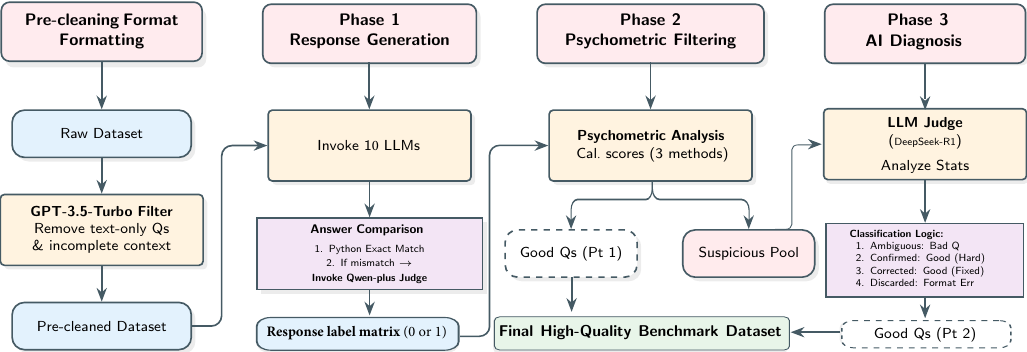}
\caption{Psychometric data cleaning pipeline. Pre-cleaning removes clearly non-computational items. Phase~1 collects multi-model responses and performs hierarchical grading. Phase~2 applies three psychometric filters to flag weak items. Phase~3 uses an AI auditor for diagnosis and correction with human evaluation.}
\label{fig:data_cleaning}
\end{figure*}

As illustrated in Fig.~\ref{fig:data_cleaning}, our cleaning pipeline adopts a funnel-style architecture comprising a rule-based pre-cleaning stage followed by three progressively deeper phases. The core insight, inspired by psychometric test theory~\cite{truong2025fantastic}, is to \emph{reverse} the evaluation direction: instead of using items to assess models. We leverage the response patterns of models with diverse proficiency levels to diagnose item quality.

\textbf{Pre-cleaning.}
A lightweight rule-based filter combined with a cost-efficient LLM (GPT-3.5-Turbo) removes clearly non-computational entries, pure essay questions, incomplete context, and items that cannot yield a quantitative answer, before entering the expensive multi-model testing phase. This initial stage prioritizes recall over precision.

\textbf{Phase~1: Multi-model response generation and hierarchical grading.}
Each surviving candidate item is submitted to a panel of heterogeneous LLMs spanning different architectures and capability tiers (e.g., GPT-4o, DeepSeek-V3, Claude-Sonnet, Qwen-Max, Gemini). A five-level hierarchical judge then grades each response in cascade: \emph{(i)}~JSON format validity, \emph{(ii)}~numerical parsing with automatic unit conversion (e.g., $10\,\text{kHz}=10{,}000\,\text{Hz}$) and a $<$1\% relative-error tolerance, \emph{(iii)}~standardized string matching after stripping LaTeX and whitespace, \emph{(iv)}~symbolic equivalence via SymPy simplification, and \emph{(v)}~LLM-as-judge semantic verification for residual ambiguous cases. This hierarchical design reduces token cost by $\sim$50\% compared with invoking an LLM judge for every instance. The output is a binary \emph{examinee$\times$item} response matrix that feeds Phase~2.

\textbf{Phase~2: Psychometric filtering.}
Three complementary metrics, each targeting a distinct quality dimension, are computed from the response matrix:
\begin{itemize}[leftmargin=*]
  \item \textbf{Metric-1: Item-Total Correlation (ITC).} We compute the point-biserial correlation between each item's binary score vector and the rest score to quantify \emph{discrimination}, i.e., the degree to which an item separates strong from weak models. Items with $r_{\text{it}}<0.15$ are flagged.

  \item \textbf{Metric-2: Mokken Scale Analysis (Loevinger's $H$).} An efficient discrimination-index proxy compares the pass rates of the top-30\% and bottom-30\% models ($\bar{p}_{\text{top}}-\bar{p}_{\text{bot}}$) to verify \emph{monotonicity}, where stronger models should not fail items that weaker models solve. Items with $H<0.30$ are flagged.

  \item \textbf{Metric-3: Inter-Item Consistency.} The mean pairwise Phi coefficient between an item and all other items measures \emph{unidimensionality}---whether the item tests the same latent trait (wireless-domain knowledge) as its peers. Items with mean $\phi<0.10$ are flagged as domain outliers.
\end{itemize}
A hierarchical filtering protocol applies these metrics in cascade: items answered correctly by all models ($p{=}1$) are exempted as \emph{anchor data} (zero variance precludes correlation checks); items with $p{=}0$ are routed directly to Phase~3 as potential ground-truth errors; ambiguous items (low pass rate with high answer-diversity entropy $>0.70$) are flagged separately; and the remaining mixed-performance items must satisfy all three thresholds to be retained.

\textbf{Phase~3: AI-assisted diagnosis and correction.}
Items entering the suspicious pool are diagnosed by a strong reasoning model acting as an auditor. For each flagged item, the auditor receives the question, the standard answer, and sampled incorrect model responses as ``error evidence.'' It then classifies the item into one of four categories: \emph{Corrected}~(ground-truth error detected and fixed), \emph{Confirmed}~(ground truth verified correct; models simply lack capability), \emph{Ambiguous}~(missing context or ill-defined conditions), or \emph{Discarded}~(non-computational or proof-based). Only \emph{Corrected} and \emph{Confirmed} items re-enter the dataset, the former with updated reference answers. Human reviewers perform a final accept/revise/reject pass to maintain editorial control. The complete mathematical formulations for all three psychometric metrics and the detailed hierarchical filtering protocol are provided in Appendix~\ref{app:psychometric}.

\subsection{Data Augmentation via Knowledge-Anchored Generation}
After cleaning, we expand each task through controlled generation strategies: parameter perturbation (\textit{e.g.}, changing SNR from 10\,dB to 15\,dB while preserving problem structure), inverse-problem templates (\textit{e.g.}, ``given rate, find bandwidth'' instead of ``given bandwidth, find rate''), difficulty scaling (adding constraints or removing simplifying assumptions), and cross-topic composition. Critically, augmented items inherit \emph{domain anchors}, such as formulas, 3GPP constraints, CQI tables, and trajectory state vectors, preventing semantically shallow paraphrase-only growth. This knowledge-anchored strategy is essential because generic augmentation techniques easily distort technical terms or destroy the logical structure of engineering problems. The full three-stage augmentation pipeline, including the knowledge-point distribution and quality verification procedure, is detailed in Appendix~\ref{app:augmentation}.

\begin{figure}[t]
\centering
\includegraphics[width=0.9\columnwidth]{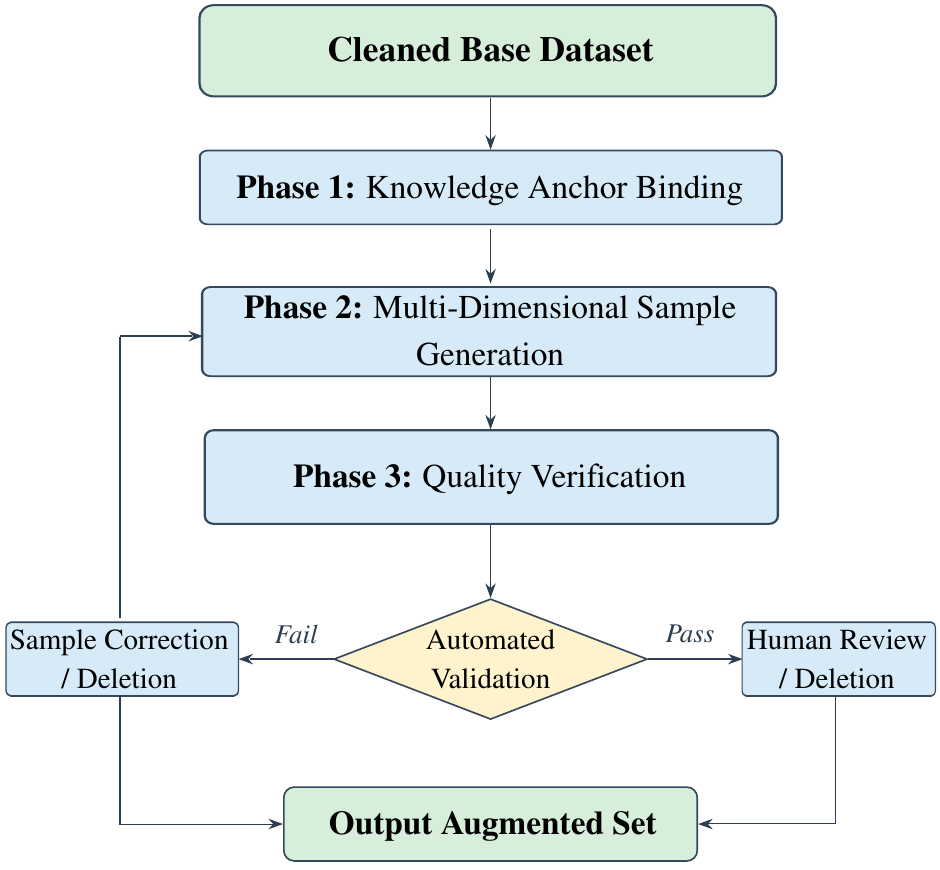}
\caption{Knowledge-anchored data augmentation pipeline. Phase~1 binds each seed item to domain anchors. Phase~2 generates multi-dimensional variants via parameter perturbation, inverse-problem templates, difficulty scaling, and cross-topic composition. Phase~3 performs automated quality verification; items that fail are corrected or discarded, with a feedback loop back to Phase~2.}
\label{fig:data_aug}
\end{figure}

\subsection{Quality Assurance: Human Validation Study}
We perform stratified human validation over all three tasks. Annotators verify question clarity, answer correctness, and rubric consistency. Inter-annotator agreement and disagreement categories are tracked to audit benchmark reliability and identify systematic annotation drift (details see Appendix~\ref{app:human}).

\subsection{Dataset Statistics and Distribution Analysis}
\begin{figure*}[t]
\centering
\includegraphics[width=0.85\linewidth]{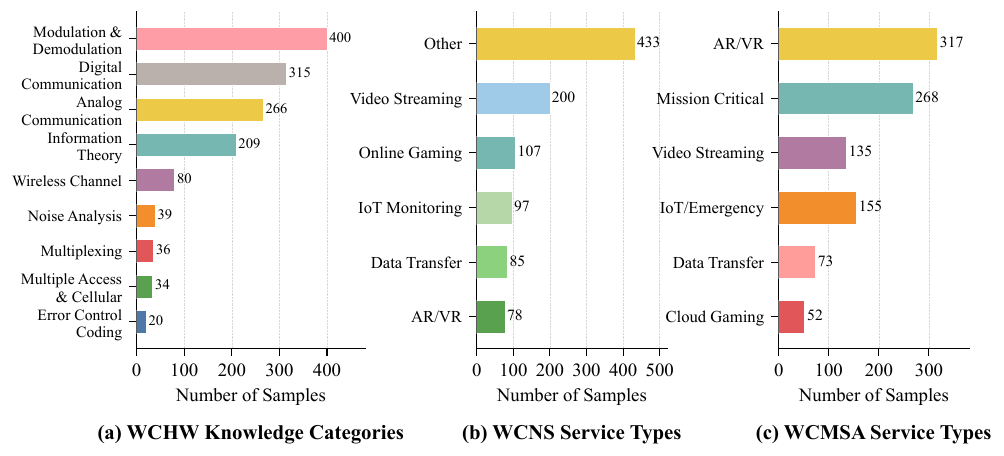}
\caption{Dataset composition of WirelessBench. (a)~WCHW knowledge category distribution across nine wireless communication topics (1,392 samples). (b)~WCNS service type distribution showing eMBB and URLLC scenario coverage (1,000 samples). (c)~WCMSA service type distribution reflecting diverse mobility-aware application scenarios (1,000 samples).}
\label{fig:dataset_composition}
\end{figure*}

The released benchmark contains 3,392 items: WCHW (1,392; test 1,044 + validation 348), WCNS (1,000; test 750 + validation 250), and WCMSA (1,000; test 750 + validation 250), with fixed validation/test splits (25\%/75\%) and deterministic references. We report distributions across knowledge topics (WCHW: 9 categories, with modulation/demodulation and digital communication as the largest groups), service intents (WCNS: eMBB $\sim$62\%, URLLC $\sim$38\%), mobility regimes (WCMSA: pedestrian 35\%, vehicular 45\%, mixed/stationary 20\%), CQI levels (1--15, ray-tracing-derived), and output types (WCHW: 78\% numeric, 15\% formula, 7\% text; WCNS/WCMSA: 100\% structured multi-field). These distributions enable reproducible subgroup analysis and targeted diagnosis of model weaknesses. Fig.~\ref{fig:dataset_composition} visualizes the detailed category and service-type distributions for all three tasks. A complete data-flow table tracing seed problems through cleaning, augmentation, and final splits is given in Appendix~\ref{app:data_flow}; representative dataset examples are provided in Appendix~\ref{app:examples}.

\begin{figure*}[t]
\centering
\includegraphics[width=0.88\linewidth]{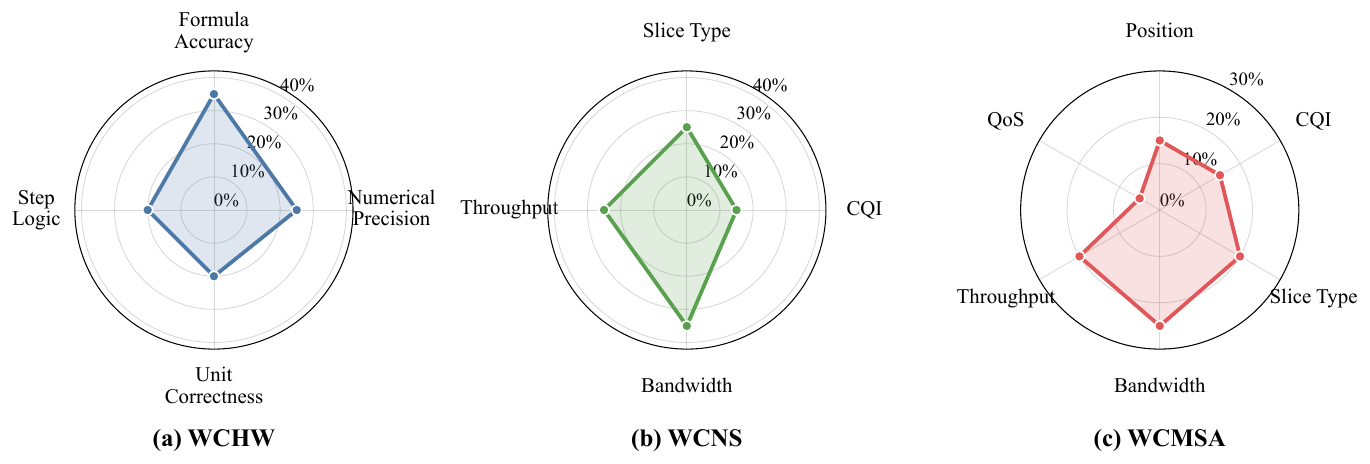}
\caption{Field-level scoring weight distributions. (a)~WCHW evaluates four dimensions: formula accuracy (35\%), numerical precision (25\%), unit correctness (20\%), and step logic (20\%). (b)~WCNS uses four-field composite scoring prioritizing bandwidth allocation (35\%) and slice type classification (25\%). (c)~WCMSA uses six-field composite scoring covering the full proactive decision chain from position prediction to QoS verification.}
\label{fig:scoring_weights}
\end{figure*}

\section{Benchmark Difficulty Calibration}
\label{sec:calibration}
After data construction, we now turn to a complementary but distinct question: \emph{how hard are the retained items, and does their difficulty structure faithfully reflect the intended three-tier cognitive hierarchy?} Psychometric filtering ensures that each item discriminates reliably among models, but it does not directly control whether the overall benchmark provides adequate coverage across difficulty strata. Therefore, we introduce a scoring-rubric design rationale, a difficulty estimation framework, per-topic and cross-task difficulty analyses, and benchmark reliability metrics. Together, these analyses provide the empirical foundation for interpreting the evaluation results reported in Section~\ref{sec:experiments}.

\subsection{Scoring Rubric Design}
\label{sec:rubric}
We first define \emph{how} each response is scored.
Section~\ref{sec:design} specifies the tolerance-aware scoring formulas (Eqs.~1--5) and per-field weight distributions (Fig.~\ref{fig:scoring_weights}).
This subsection complements that design by detailing the \emph{implementation-level} grading pipeline that translates raw model outputs into scores.

\textbf{Answer-type routing.}
The evaluator first classifies each reference answer into one of four types:
(i)~\emph{numeric / scientific notation} (e.g., $6.87$, $2.13\!\times\!10^{-2}$, \texttt{1.5e-6});
(ii)~\emph{formula / symbolic expression} (e.g., $C\!=\!B\log_2(1\!+\!\text{SNR})$);
(iii)~\emph{short technical text} (e.g., ``BPSK'', ``Rayleigh fading'');
(iv)~\emph{structured multi-field output} (JSON-like records for WCNS / WCMSA).

\textbf{Numeric scoring details.}
We parse scientific notation in multiple formats, i.e., standard ($1.5\!\times\!10^{-6}$), programming (\texttt{1.5e-6}), to handle diverse model output styles.
Before computing the relative error (Eq.~1), the evaluator normalizes units to base SI using the conversion rules in Table~\ref{tab:unit_conversion}.

\begin{table}[t]
\centering
\caption{Supported unit families and conversion factors.}
\label{tab:unit_conversion}
\footnotesize
\begin{tabular}{@{}lp{4.4cm}c@{}}
\toprule
\textbf{Family} & \textbf{Units} & \textbf{Base} \\
\midrule
Frequency & Hz, kHz ($\!\times\!10^3$), MHz ($\!\times\!10^6$), GHz ($\!\times\!10^9$) & Hz \\
Power (lin.) & $\mu$W ($\!\times\!10^{-6}$), mW ($\!\times\!10^{-3}$), W & W \\
Power (log) & dBm, dBW ($P_{\text{dBW}}\!=\!P_{\text{dBm}}\!-\!30$) & dBm \\
Data rate & bps, kbps ($\!\times\!10^3$), Mbps ($\!\times\!10^6$), Gbps ($\!\times\!10^9$) & bps \\
Distance & m, km ($\!\times\!10^3$) & m \\
\bottomrule
\end{tabular}
\end{table}

\textbf{Catastrophic error detection.}
Two error categories receive an automatic zero score regardless of numeric proximity:
(i)~\emph{unit mismatch}: for example, ``30~dBm'' versus the reference ``30~dB'' is penalized to zero because 30~dBm~$=1$~W whereas 30~dB is a dimensionless ratio of 1{,}000;
(ii)~\emph{order-of-magnitude error}: when the predicted exponent disagrees with the reference exponent by $\ge1$ (e.g., $10^{-3}$ vs.\ $10^{-6}$).
These checks initialize before tolerance-tier scoring.

\textbf{Formula and text scoring.}
Formula answers undergo normalization and are compared via character-level edit distance combined with structural similarity matching; full credit requires symbolic equivalence, and partial credit is awarded for correct structure with minor notational differences.
Text answers are evaluated by weighted keyword overlap and semantic consistency.

\textbf{Structured-output grading.}
For WCNS and WCMSA, each field is scored independently by its type-specific sub-evaluator, then aggregated by the weighted sums defined in Eqs.~3--4.
Slice type uses exact-match; CQI additionally checks CQI$\in[1,15]$; bandwidth and throughput include unit-consistency verification (e.g., rejecting kbps when the ground truth is in Mbps).

\textbf{Rubric calibration procedure.}
The field weights were established through a three-round iterative process:
(i)~\emph{initial proposal}, where domain experts assigned weights based on operational-impact analysis;
(ii)~\emph{pilot grading} on a 200-item subset across five LLMs to verify meaningful score variance;
(iii)~\emph{weight adjustment} for fields that caused score compression or inversion.
For example, CQI weight in WCNS was initially 0.25 but reduced to 0.15 after pilot results showed CQI variance was dominated by binary tool-call behavior rather than reasoning quality.

\subsection{Difficulty Estimation Methodology}
We estimate item difficulty using a composite index that combines three normalized components: (i)~$r_i$, the number of required reasoning steps (normalized by the maximum across the task); (ii)~$o_i$, arithmetic and operation complexity (number of distinct operations such as log, division, unit conversion); and (iii)~$m_i$, cross-domain integration burden (number of distinct knowledge domains or tool outputs the item requires). The composite difficulty index is:
\begin{equation}
D_i = \alpha\, r_i + \beta\, o_i + \gamma\, m_i,
\end{equation}
where $\alpha+\beta+\gamma=1$. For WCHW, we set $\alpha\!=\!0.4,\,\beta\!=\!0.4,\,\gamma\!=\!0.2$ (reasoning and computation dominate, cross-domain integration is minimal). For WCNS, we use $\alpha\!=\!0.3,\,\beta\!=\!0.3,\,\gamma\!=\!0.4$ (tool integration raises the cross-domain burden). For WCMSA, $\alpha\!=\!0.35,\,\beta\!=\!0.25,\,\gamma\!=\!0.40$ (multi-step dependence and tool chaining are primary difficulty drivers). Items are then binned into three strata: \emph{easy} ($D_i < 0.33$), \emph{medium} ($0.33 \le D_i < 0.66$), and \emph{hard} ($D_i \ge 0.66$).

To validate the composite index empirically, we compute the Pearson correlation between $D_i$ and the observed item error rate (fraction of models answering incorrectly). Across WCHW, WCNS, and WCMSA, the correlations are $0.61$, $0.58$, and $0.67$, respectively ($p < 0.001$ in all cases), confirming that the index captures meaningful variation in item difficulty.

\subsection{Per-Topic Difficulty Distribution (WCHW)}
For WCHW, we report category-level difficulty distributions across the nine knowledge categories. Topics involving multi-step derivations (e.g., coding-theoretic channel capacity, OFDM subcarrier allocation) exhibit disproportionate hard-item concentration ($>$45\% hard), whereas direct formula-substitution topics (e.g., free-space path loss, thermal noise power) are dominated by easy/medium items ($>$70\% easy + medium). This asymmetry is intentional: it helps distinguish ``knowledge coverage gaps'' (the model lacks the right formula) from ``reasoning-depth gaps'' (the model knows the formula but cannot chain multi-step derivations).

\subsection{Cross-Task Difficulty Comparison}
We compare the difficulty profile across the three tiers to verify hierarchy consistency. WCHW's difficulty is predominantly reasoning-driven (high $r_i$, moderate $o_i$, low $m_i$); WCNS shifts the difficulty source toward cross-domain integration ($m_i$), because the agent must bridge intent understanding, tool invocation, and allocation computation; WCMSA further increases both $r_i$ and $m_i$, as six-field decisions require longer reasoning chains with mandatory tool chaining. The median $D_i$ increases monotonically across tiers (WCHW: 0.38, WCNS: 0.49, WCMSA: 0.57), confirming that the cognitive hierarchy is reflected in empirical difficulty. This monotonic progression ensures that aggregate model scores can be meaningfully decomposed by tier and that strong Tier-1 performance is a necessary but not sufficient condition for strong Tier-2/3 performance.




\section{Experimental Evaluation}
\label{sec:experiments}

This section evaluates \wb{} from three complementary perspectives: aggregate performance across model families and method categories, metric sensitivity (exact match vs.\ tolerance-aware), and failure-mode diagnostics.

\subsection{Experimental Setup}
\label{sec:setup}

\textbf{Models and baselines.} We evaluate four categories of approaches. (i)~\emph{Direct prompting:} Qwen-Turbo-Latest~\cite{qwen2024qwen25} and GPT-4o~\cite{openai2024gpt4o} with zero-shot prompts. (ii)~\emph{Advanced prompting:} Chain-of-Thought Self-Consistency (CoT-SC, $k\!=\!5$)~\cite{wei2022cot,wang2023selfconsistency} and MedPrompt~\cite{nori2023medprompt} with GPT-4o. (iii)~\emph{Automated agentic methods:} ADAS~\cite{hu2024adas} and AFlow~\cite{zhang2025aflow}. (iv)~\emph{Task-aware reference pipeline (\wbr{}):} AFlow and \wbr{} both augmented with wireless-specific tools (Python executor, unit converter, formula verifier, ray-tracing predictor, Kalman filter), detailed in~\cite{tong2025wirelessagentpp} and Appendix~\ref{app:ref_agent}.

\textbf{Protocol and metrics.} All methods share a unified prompting template, fixed decoding temperature ($T\!=\!0$), deterministic post-processing, and the tolerance-aware scoring pipeline (Appendix~\ref{app:scoring}). We report per-task scores (WCHW accuracy, WCNS/WCMSA composite), macro-average benchmark score, per-field scores for structured tasks, and 95\% bootstrap confidence intervals (1,000 resamples). Model specifications and hyperparameters are listed in Appendix~\ref{app:exp_config}.

\subsection{Main Results}
\label{sec:main_results}

Table~\ref{tab:leaderboard_main} reports the primary leaderboard. We summarize three key findings from these results as follows.

\textbf{Finding~1: Direct prompting is far from operationally sufficient.} GPT-4o achieves only $68.00\%$ on average, with WCHW being the hardest tier ($60.32\%$) due to multi-step symbolic reasoning involving transcendental functions and unit conversions. Qwen-Turbo-Latest reaches $62.30\%$, confirming that even frontier models struggle without tool support.

\textbf{Finding~2: Prompting improvements yield diminishing returns.} CoT-SC and MedPrompt improve over direct GPT-4o prompting by only $1.46$\,pp and $1.10$\,pp on average. This plateau suggests that the bottleneck is \emph{reliable numerical computation and tool orchestration}, not reasoning elicitation.

\textbf{Finding~3: Tool access provides a large performance uplift, but part of the gap reflects information asymmetry.} AFlow reaches $73.29\%$ and \wbr{} achieves $84.64\%$, where a $16.64$\,pp gap over the best direct prompting score. However, this comparison conflates two effects: (a)~the ability to orchestrate tools effectively, and (b)~access to information (CQI, predicted position) that direct-prompting methods structurally cannot obtain. On the WCNS CQI field, direct-prompting methods score $<5\%$ because they must guess a value that is only available via the ray-tracing tool. The AFlow-vs-GPT-4o gap ($+5.29$\,pp), where both lack co-design coupling, provides a cleaner (albeit partial) estimate of agentic workflow benefit. Future Oracle experiments (providing CQI and/or position directly) are needed to fully decompose the information-access and tool-orchestration contributions.

\begin{table}[t]
\centering
\caption{Main leaderboard on WirelessBench test set. Scores are tolerance-aware (\%). Best result per column in \textbf{bold}; best non-agent result \underline{underlined}.}
\label{tab:leaderboard_main}
\resizebox{\columnwidth}{!}{
\begin{tabular}{llcccc}
\toprule
\textbf{Category} & \textbf{Method} & \textbf{WCHW} & \textbf{WCNS} & \textbf{WCMSA} & \textbf{Avg.} \\
\midrule
\multirow{2}{*}{Direct Prompt}
 & Qwen-Turbo-Latest~\cite{qwen2024qwen25} & 58.34 & 62.13 & 66.43 & 62.30 \\
 & GPT-4o~\cite{openai2024gpt4o} & 60.32 & \underline{72.45} & 71.22 & 68.00 \\
\midrule
\multirow{2}{*}{Adv.\ Prompting}
 & CoT-SC ($k\!=\!5$)~\cite{wei2022cot} & 60.01 & 74.82 & \underline{73.56} & 69.46 \\
 & MedPrompt~\cite{nori2023medprompt} & \underline{61.22} & 73.18 & 72.89 & \underline{69.10} \\
\midrule
\multirow{2}{*}{Agentic}
 & ADAS~\cite{hu2024adas} & 53.13 & 68.42 & 65.41 & 62.32 \\
 & AFlow~\cite{zhang2025aflow} & 69.92 & 76.12 & 73.90 & 73.29 \\
\midrule
Reference Pipeline
 & \wbr{}~\cite{tong2025wirelessagentpp} & \textbf{81.02} & \textbf{86.18} & \textbf{86.72} & \textbf{84.64} \\
\bottomrule
\end{tabular}}
\end{table}

\subsection{Metric Sensitivity: Exact Match vs.\ Tolerance-Aware}
\label{sec:sensitivity}

Table~\ref{tab:sensitivity_rank} compares tolerance-aware and strict exact-match scoring (relative error $\leq 0.1\%$). Under exact match, all methods suffer score drops, but the drops are highly uneven: GPT-4o loses $16.18$\,pp ($68.00\% \to 51.82\%$) because many of its numerical answers are close but not exact, while \wbr{} loses only $5.59$\,pp ($84.64\% \to 79.05\%$) because tool-verified answers are predominantly exact. The key diagnostic insight is that exact-match scoring conflates two failure modes: minor numerical imprecision (benign, tolerance-rescuable) and catastrophic unit/magnitude errors (deployment-dangerous, not rescuable). Tolerance-aware scoring separates these modes, providing a more faithful risk assessment. We note that this comparison examines only one axis of sensitivity (exact vs.\ tolerance-aware); a full sweep over threshold values (e.g., 0.1\%/1\%/5\%/10\%/20\%) and field-weight perturbations is planned as future work to confirm ranking robustness.

\begin{table}[t]
\centering
\caption{Ranking sensitivity under exact-match vs.\ tolerance-aware scoring. $S$: macro-average (\%); $R$: rank (1=best). Qwen-Turbo-Latest and MedPrompt are omitted; their inclusion does not alter the top-3 ranking.}
\label{tab:sensitivity_rank}
\begin{tabular}{lcccccc}
\toprule
\textbf{Method} & $S^{\text{exact}}$ & $R^{\text{exact}}$ & $S^{\text{tol}}$ & $R^{\text{tol}}$ & $\Delta R$ \\
\midrule
GPT-4o & 51.82 & 5 & 68.00 & 5 & $0$ \\
CoT-SC ($k\!=\!5$) & 53.14 & 4 & 69.46 & 3 & $+1$ \\
ADAS & 52.07 & 6 & 62.32 & 6 & $0$ \\
AFlow & 66.61 & 2 & 73.29 & 2 & $0$ \\
\wbr{} & 79.05 & 1 & 84.64 & 1 & $0$ \\
\bottomrule
\end{tabular}
\end{table}

\subsection{Failure-Mode Decomposition}
\label{sec:failure_mode}

\begin{figure}[t]
\centering
\includegraphics[width=\columnwidth]{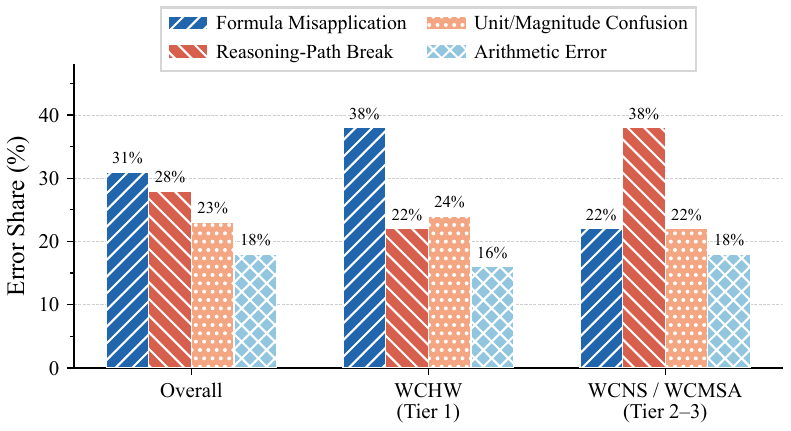}
\caption{Failure-mode decomposition of GPT-4o (direct prompting) errors across benchmark tiers. Formula misapplication dominates on WCHW (Tier~1), while reasoning-path breaks dominate on WCNS/WCMSA (Tiers~2--3). Unit/magnitude confusion is uniformly distributed across all tiers.}
\label{fig:failure_mode}
\end{figure}

We tag all mispredicted samples from GPT-4o (direct prompting) into four error categories:
\begin{itemize}[leftmargin=*]
  \item \textbf{Formula misapplication (31\%):} Wrong equation or incorrect variable substitution (e.g., coherent vs.\ non-coherent BFSK formula, linear vs.\ dB-scale SNR).
  \item \textbf{Reasoning-path break (28\%):} Correct intermediate steps but failure at a chain junction (e.g., computing bandwidth correctly but using it as throughput).
  \item \textbf{Unit/magnitude confusion (23\%):} dB/dBm, MHz/Hz, or mW/W mixups causing order-of-magnitude errors---the most deployment-dangerous category.
  \item \textbf{Arithmetic error (18\%):} Correct reasoning but computation mistakes on scientific notation or transcendental functions.
\end{itemize}
Fig.~\ref{fig:failure_mode} visualizes the error distribution across tiers. On WCHW, formula misapplication dominates ($38\%$); on WCNS/WCMSA, reasoning-path breaks lead ($35$--$40\%$). Unit/magnitude errors are uniformly distributed ($20$--$25\%$) across tiers but nearly absent from tool-using agents, confirming that tool-integrated verification is an effective safeguard against catastrophic failures.

\section{Discussion}
\label{sec:discussion}

\textbf{Co-design caveat.}
\label{sec:co_design}
Because \wbr{} was built by the same team and shares the tool interface, field structure, and scoring rubric of WirelessBench, its high score may partly reflect \emph{benchmark--system co-design} rather than generalizable superiority; we therefore position it as a \emph{diagnostic ceiling} rather than an unbiased upper bound. AFlow and ADAS, two third-party agentic frameworks with no co-design relationship, also outperform direct-prompting methods on certain tiers, providing independent evidence that agentic workflow structure itself improves performance. Nevertheless, Oracle ablations (providing CQI/position directly to non-tool methods), third-party tool-agent baselines, and same-backbone progressive-access experiments are needed to fully decompose information-access from tool-orchestration contributions; we plan to include them in the next release.

\textbf{Threats to validity and limitations.}
The current digital twin covers a single campus (HKUST, 3.5\,GHz n78, 3GPP UMi), so absolute scores may not generalize to other deployments; relative rankings among methods are expected to be more portable but require cross-environment verification. The baseline set relies on two closed-source LLMs and lacks open-source and domain-adapted models. Scoring thresholds and field weights are engineering heuristics; partial evidence of rank stability is shown in Section~\ref{sec:sensitivity}, but comprehensive sensitivity sweeps are planned. Approximately 74\% of items originate from LLM-assisted augmentation, raising potential diversity and style-leakage concerns; seed-vs.-augmented subset analysis and embedding-based diversity audits are planned. WCMSA's constant-velocity Kalman filter oversimplifies real mobility, and all tasks assume a single agent with static, offline data. Extending the benchmark to heterogeneous environments, multi-agent and streaming settings, open-source baselines, human-expert anchoring, and stricter deduplication constitutes our ongoing work. WirelessBench is designed for evaluation only; any deployment-facing use of LLM outputs requires human oversight and fail-safe fallbacks.

\section{Conclusions}
\label{sec:conclusion}
We presented \wb{}, an extensible benchmark for evaluating the operational reliability of LLM agents on wireless network intelligence. \wb{} emerged as the principal deployment bottleneck
during the development of wireless AI agents.
Built around a three-tier cognitive hierarchy (WCHW, WCNS, WCMSA) that spans static domain knowledge, intent-to-allocation mapping, and
proactive mobility-aware service assurance, the benchmark operationalizes reliability through three design commitments: tolerance-aware scoring, tool-necessary tasks, and per-item CoT. Evaluating seven methods reveals that the strongest frontier model achieves only $68.00\%$ under
direct prompting, and that formula misapplication ($31\%$), reasoning-path breaks ($28\%$), and unit confusion ($23\%$) dominate the failure landscape (the last of which is nearly eliminated by tool-integrated verification). While current limitations in environment coverage, data augmentation, and baseline diversity remain open, we hope \wb{} provides a rigorous and community-accessible foundation for advancing the reliability of AI agents in mobile-network operations. The code and data are available at \url{https://github.com/jwentong/WirelessBench}

\bibliographystyle{IEEEtran}
\bibliography{WB_Ref}

\appendices

\section{Scoring Protocol Summary}\label{app:scoring}

The complete scoring pipeline---answer-type routing, unit-conversion rules, tolerance-aware numeric scoring, catastrophic-error detection, formula/text scoring, and structured-output field-level aggregation---is described in Section~\ref{sec:rubric}. Table~\ref{tab:field_weights} below provides a consolidated reference of the field-level weights.

\begin{table}[h]
\centering
\caption{Field-level scoring weights for structured tasks.}
\label{tab:field_weights}
\footnotesize
\begin{tabular}{@{}lcc@{}}
\toprule
\textbf{Field} & \textbf{WCNS Weight} & \textbf{WCMSA Weight} \\
\midrule
Position Prediction & --- & 15\% \\
CQI Prediction & 15\% & 15\% \\
Slice Type Classification & 25\% & 20\% \\
Bandwidth Allocation & 35\% & 25\% \\
Throughput Estimation & 25\% & 20\% \\
QoS Satisfaction & --- & 5\% \\
\bottomrule
\end{tabular}
\end{table}

\section{3GPP CQI-to-Spectral-Efficiency Mapping}\label{app:cqi}

Table~\ref{tab:cqi} presents the standardized CQI-to-spectral-efficiency mapping from 3GPP TS~38.214 Table~5.2.2.1-3, which is used for all throughput calculations in WCNS and WCMSA tasks.

\begin{table}[h]
\centering
\caption{3GPP TS~38.214 CQI Table (4-bit CQI, Table~5.2.2.1-3).}
\label{tab:cqi}
\footnotesize
\begin{tabular}{@{}cccc@{}}
\toprule
\textbf{CQI} & \textbf{Modulation} & \textbf{Code Rate $\times 1024$} & \textbf{Spectral Eff.\ (bps/Hz)} \\
\midrule
1 & QPSK & 78 & 0.15 \\
2 & QPSK & 120 & 0.23 \\
3 & QPSK & 193 & 0.38 \\
4 & QPSK & 308 & 0.60 \\
5 & QPSK & 449 & 0.88 \\
6 & QPSK & 602 & 1.18 \\
7 & 16QAM & 378 & 1.48 \\
8 & 16QAM & 490 & 1.91 \\
9 & 16QAM & 616 & 2.41 \\
10 & 64QAM & 466 & 2.73 \\
11 & 64QAM & 567 & 3.32 \\
12 & 64QAM & 666 & 3.90 \\
13 & 64QAM & 772 & 4.52 \\
14 & 64QAM & 873 & 5.12 \\
15 & 64QAM & 948 & 5.55 \\
\bottomrule
\end{tabular}
\end{table}

\textbf{Throughput Calculation.} Given bandwidth $B$ (MHz) and CQI index, the achievable throughput is:
\begin{equation}
R = B \times \eta_{\text{CQI}} \quad (\text{Mbps}),
\end{equation}
where $\eta_{\text{CQI}}$ is the spectral efficiency from Table~\ref{tab:cqi}.

\textbf{Example.} For CQI~$=8$ and $B=6.92$~MHz: $R = 6.92 \times 1.91 = 13.2$~Mbps.

\section{Digital-Twin and Ray-Tracing Configuration}\label{app:digital_twin}

WCNS and WCMSA tasks are grounded in a campus-scale digital twin of the Hong Kong University of Science and Technology (HKUST) campus. This section documents the ray-tracing setup and propagation modeling.

\subsection{Geographic Coverage}

The digital twin covers three distinct regions of the HKUST campus:
\begin{itemize}[leftmargin=*]
  \item \textbf{North Region:} Academic buildings and pedestrian corridors. Characterized by dense indoor--outdoor transitions and multipath-rich propagation.
  \item \textbf{Center Region:} Open plaza and mixed-use areas. Relatively clear line-of-sight (LOS) conditions with moderate shadowing.
  \item \textbf{South Region:} Residential halls and hillside terrain. Significant elevation variation and NLOS-dominated propagation.
\end{itemize}

\subsection{Ray-Tracing Parameters}

Propagation data is pre-computed using ray-tracing simulation conforming to the 3GPP TR~38.901 Urban Micro (UMi) channel model:
\begin{itemize}[leftmargin=*]
  \item \textbf{Carrier Frequency:} 3.5~GHz (n78 band)
  \item \textbf{Subcarrier Spacing:} 30~kHz
  \item \textbf{Base Station Tx Power:} 46~dBm
  \item \textbf{Base Station Height:} 10~m
  \item \textbf{User Equipment Height:} 1.5~m
  \item \textbf{Path Loss Model:} 3GPP TR~38.901 UMi-Street Canyon
    \begin{itemize}[nosep]
      \item LOS: $PL = 32.4 + 21\log_{10}(d) + 20\log_{10}(f_c)$
      \item NLOS: $PL = 35.3\log_{10}(d) + 22.4 + 21.3\log_{10}(f_c) - 0.3(h_{UE}-1.5)$
    \end{itemize}
  \item \textbf{Shadow Fading:} Log-normal with $\sigma_{\text{LOS}}=4$~dB, $\sigma_{\text{NLOS}}=7.82$~dB
  \item \textbf{LOS Probability:} $\Pr(\text{LOS}) = \min(18/d, 1)(1 - e^{-d/36}) + e^{-d/36}$
\end{itemize}

\subsection{SNR-to-CQI Mapping}

The received SNR at each user position is mapped to CQI index (1--15) using the 3GPP-defined thresholds. The mapping ensures that CQI = 15 corresponds to the highest spectral efficiency ($\eta = 5.55$~bps/Hz) at SNR~$\geq 19.76$~dB, while CQI = 1 corresponds to the lowest ($\eta = 0.15$~bps/Hz) at SNR~$\in [-6.7, -4.4]$~dB.

\subsection{Tool Interface}

The ray-tracing tool is invoked via a function-call interface:
\begin{tcolorbox}[colback=gray!5!white, colframe=gray!60!black]
\footnotesize
\texttt{cqi = ray\_tracing(x=\textit{float}, y=\textit{float}, region=\textit{str})}
\\[2pt]
\textbf{Inputs:} $(x,y)$ user coordinates in meters; \texttt{region} $\in$ \{\texttt{"North"}, \texttt{"Center"}, \texttt{"South"}\}.
\\
\textbf{Output:} Integer CQI $\in [1, 15]$.
\end{tcolorbox}

Agents must correctly parse user coordinates from the problem text, identify the geographic region, and supply well-formed arguments. The tool returns a deterministic CQI value for each $(x,y,\text{region})$ tuple, enabling reproducible evaluation.

\section{Psychometric Data-Cleaning Pipeline}\label{app:psychometric}

Raw datasets in the wireless communication domain suffer from several quality issues: (1)~non-computational noise (pure concept explanations or essay-style questions), (2)~incorrect ground truth (errors in original question banks), and (3)~ambiguity (poorly defined problem descriptions). We design a multi-stage ``funnel-style'' cleaning pipeline to address these issues.

\subsection{Pre-Cleaning: Rule-Based Filtering}

Before the computationally expensive full-model testing phase, we perform preliminary noise reduction:
\begin{itemize}[leftmargin=*]
  \item \textbf{Hard Rules:} Directly discard pure essay questions containing keywords like ``explain the principle'' or ``discuss the advantages'' where the answer contains no mathematical formulas.
  \item \textbf{Soft Filter:} For ambiguous questions, a lightweight LLM determines whether each question requires a specific numerical value, formula, or structured output.
\end{itemize}

\subsection{Phase~1: Response Generation and Hierarchical Grading}

Cleaned questions are input into 10 representative LLMs (including Qwen-Max, DeepSeek-V3, GPT-4o, Claude-3.5-Sonnet) to construct an ``examinee--item'' response matrix $\mathbf{X} \in \{0,1\}^{N \times M}$, where $N$ is the number of items and $M=10$ is the number of models. We implement a five-level hierarchical grading system:
\begin{enumerate}[leftmargin=*]
  \item \textbf{Level~1 (Format Check):} Validates JSON integrity and required field presence.
  \item \textbf{Level~2 (Metric Parsing):} Parses numerical values with units (e.g., $10\text{k}=10{,}000$, $10\,\text{mW}=0.01$\,W) and computes relative error $|e|<1\%$.
  \item \textbf{Level~3 (String Matching):} Compares standardized formula strings after normalization.
  \item \textbf{Level~4 (SymPy Verification):} Converts answers into symbolic expressions via SymPy and verifies mathematical equivalence.
  \item \textbf{Level~5 (LLM Judge):} Only when all above rules fail, invokes an advanced LLM for semantic judgment.
\end{enumerate}
This hierarchical structure reduces token cost by $\sim$50\% compared with directly invoking an LLM judge for every instance.

\subsection{Phase~2: Psychometric Filtering Framework}

Inspired by psychometric theory~\cite{truong2025fantastic}, we treat LLMs as ``examinees'' and evaluate item quality through three statistical metrics.

\textbf{Metric~1: Item-Total Correlation (Discrimination).}
We use the Point-Biserial Correlation with Rest Score correction:
\begin{equation}
r_{it} = \text{Pearson}(\mathbf{x}_i, \mathbf{R}_i), \quad \text{where } R_{im} = \sum_{j \neq i} x_{jm}.
\end{equation}
Items with $r_{it} < 0.15$ are flagged as having low discrimination---i.e., high- and low-performing models score roughly equally on the item, indicating that it does not differentiate capability levels.

\textbf{Metric~2: Mokken Scale Analysis (Monotonicity).}
We employ a Discrimination Index as an efficient proxy for full Mokken scalability:
\begin{equation}
\text{Mokken\_Proxy}_i = \bar{p}_{\text{top}} - \bar{p}_{\text{bottom}},
\end{equation}
where $\bar{p}_{\text{top}}$ and $\bar{p}_{\text{bottom}}$ are pass rates of the top/bottom 30\% of models. Items with Mokken\_Proxy $< 0.30$ violate monotonicity---a weaker model may paradoxically outperform a stronger one, indicating a quality issue with the item.

\textbf{Metric~3: Inter-Item Consistency (Unidimensionality).}
We measure domain alignment via mean pairwise correlation:
\begin{equation}
\text{Score}_{\text{consistency}}^{(i)} = \frac{1}{M-1} \sum_{j \neq i} \text{Corr}(\mathbf{x}_i, \mathbf{x}_j).
\end{equation}
Items with score $< 0.10$ are flagged as outliers that do not align with the overall construct.

\textbf{Hierarchical Filtering Protocol.}
We apply a cascade of filters: (1)~content validity: filter non-quantitative questions; (2)~ambiguity detection: flag items with low pass rate but high answer diversity; (3)~extreme consistency management: exempt all-pass items as ``anchor data''; (4)~psychometric audit for remaining items using the three metrics above.

\subsection{Phase~3: AI-Assisted Diagnosis and Correction}

Questions flagged in Phase~2 enter a ``suspicious pool'' for deep diagnosis using DeepSeek-R1 as an ``auditor.'' The auditor classifies each flagged item into one of four categories:
\begin{itemize}[leftmargin=*]
  \item \textbf{CORRECTED:} The original ground truth is wrong; the AI provides the correct derivation and updated answer.
  \item \textbf{CONFIRMED:} The original ground truth is correct; models failed due to difficulty.
  \item \textbf{AMBIGUOUS:} The question lacks key conditions and cannot be uniquely solved.
  \item \textbf{DISCARDED:} Confirmed as non-computational or conceptual.
\end{itemize}
This workflow not only cleans data but also automatically repairs questions with incorrect answers, providing a high-quality benchmark.

\section{Data Augmentation Pipeline}\label{app:augmentation}

Starting from 312 seed problems per task (936 total), after psychometric cleaning 888 high-quality samples remain. To achieve sufficient scale, we apply knowledge-anchored augmentation, expanding to 3,392 final samples (1,392 WCHW + 1,000 WCNS + 1,000 WCMSA). Standard augmentation techniques (back-translation, synonym replacement) are inappropriate for wireless problems because they easily distort technical terms or destroy mathematical structure. Instead, we use a three-stage framework.

\subsection{Stage~1: Knowledge Point Anchoring}

We construct a knowledge graph covering nine core wireless communication topics to ensure augmented samples remain within the target domain:

\begin{table}[h]
\centering
\caption{WCHW knowledge point distribution after augmentation. Counts reflect the augmented pool; the released set contains 1{,}392 items after deduplication.}
\label{tab:knowledge_dist}
\footnotesize
\begin{tabular}{@{}lcc@{}}
\toprule
\textbf{Knowledge Category} & \textbf{Count} & \textbf{Percentage} \\
\midrule
Modulation \& Demodulation & 400 & 28.7\% \\
Digital Communication & 315 & 22.6\% \\
Analog Communication & 266 & 19.1\% \\
Information Theory & 209 & 15.0\% \\
Wireless Channel \& Fading & 80 & 5.7\% \\
Noise Analysis & 39 & 2.8\% \\
Multiplexing (FDM/TDM/CDM) & 36 & 2.6\% \\
Multiple Access \& Cellular & 34 & 2.4\% \\
Error Control Coding & 20 & 1.4\% \\
\bottomrule
\end{tabular}
\end{table}

\subsection{Stage~2: Multi-Dimensional Sample Generation}

Using advanced LLMs with carefully designed prompts, we apply four augmentation strategies:
\begin{itemize}[leftmargin=*]
  \item \textbf{Parameter Substitution:} Systematically replace numerical values while preserving problem structure (e.g., $R_b = 64$\,kbps $\to$ $R_b = 128$\,kbps; $\alpha = 0.3 \to 0.5, 0.7, 1.0$). Logical consistency is verified after substitution.
  \item \textbf{Question Type Transformation:} Convert between equivalent question types within the same knowledge point (e.g., given bandwidth $\to$ find rate $\Leftrightarrow$ given rate $\to$ find bandwidth; given $E_b/N_0$ $\to$ find BER $\Leftrightarrow$ given BER $\to$ find $E_b/N_0$).
  \item \textbf{Difficulty Scaling:} Add constraints or remove simplifying assumptions to create harder variants.
  \item \textbf{Context Variation:} Change application scenarios while preserving core calculations (e.g., LTE $\to$ 5G NR, indoor $\to$ outdoor).
\end{itemize}

\subsection{Stage~3: Quality Verification}

Every generated sample undergoes:
\begin{enumerate}[leftmargin=*]
  \item \textbf{Format Validation:} Ensures JSON compliance with all required fields (\texttt{question}, \texttt{answer}, \texttt{cot}, \texttt{id}).
  \item \textbf{Solution Verification:} Answers are cross-checked using deterministic numerical solvers or symbolic computation (SymPy); all ground-truth answers are recomputed rather than taken from LLM output.
  \item \textbf{Deduplication:} TF-IDF-based cosine similarity is computed between all pairs; samples with similarity $>0.85$ are deduplicated.
  \item \textbf{Domain Consistency Check:} Verifies knowledge-point alignment with the source problem.
  \item \textbf{Difficulty Calibration:} Testing on multiple LLMs ensures reasonable difficulty spread.
\end{enumerate}

\section{Human Validation Protocol}\label{app:human}

\subsection{Sampling Methodology}

We perform stratified manual validation across all three tiers with balanced sampling over difficulty bins and topic groups. We sample 150 problems (50 per task) using:
\begin{itemize}[leftmargin=*]
  \item \textbf{Difficulty strata:} Easy (model pass rate $>80\%$), Medium ($40$--$80\%$), Hard ($<40\%$), with equal allocation per stratum.
  \item \textbf{Annotators:} 3 graduate students in Electrical Engineering with wireless communication coursework, each with $\geq 2$ years of research experience.
  \item \textbf{Task:} Verify (1) question clarity, (2) answer correctness, and (3) CoT validity.
\end{itemize}

\subsection{Validation Results}

\begin{table}[h]
\centering
\caption{Human validation results (150 samples, 3 annotators).}
\label{tab:human_val}
\scriptsize
\begin{tabular}{@{}lcccc@{}}
\toprule
\textbf{Metric} & \textbf{WCHW} & \textbf{WCNS} & \textbf{WCMSA} & \textbf{Overall} \\
\midrule
Question Clarity & 96\% & 98\% & 94\% & 96.0\% \\
Answer Correctness & 94\% & 98\% & 96\% & 96.0\% \\
CoT Validity & 92\% & 96\% & 90\% & 92.7\% \\
Inter-Annotator Agreement ($\kappa$) & 0.87 & 0.91 & 0.84 & 0.87 \\
\bottomrule
\end{tabular}
\end{table}

\subsection{Common Error Types Found}

Human review identified the following issues (subsequently corrected before final release):
\begin{itemize}[leftmargin=*]
  \item \textbf{WCHW (3 errors):} Rounding inconsistencies in multi-step calculations where accumulated rounding differs between intermediate steps and the final answer.
  \item \textbf{WCNS (1 error):} Ambiguous service description (``real-time gaming'' could be classified as either eMBB or URLLC depending on latency sensitivity interpretation).
  \item \textbf{WCMSA (2 errors):} Kalman prediction sign errors in velocity estimation when trajectory involves direction reversal.
\end{itemize}
The 96\% pre-correction answer correctness rate validates the effectiveness of the multi-model automated pipeline.

\section{Data Pipeline Summary}\label{app:data_flow}

Table~\ref{tab:data_flow} provides the complete data pipeline from seed problems to final benchmark splits.

\begin{table}[h]
\centering
\caption{WirelessBench data pipeline: from seed problems to final benchmark splits.}
\label{tab:data_flow}
\scriptsize
\begin{tabular}{@{}lcccccc@{}}
\toprule
\textbf{Task} & \textbf{Seed} & \textbf{Clean} & \textbf{Aug.} & \textbf{Final} & \textbf{Val/Test} & \textbf{Source} \\
\midrule
WCHW & 312 & 280 & 1,112 & 1,392 & 348/1,044 & Textbook \\
WCNS & 312 & 304 & 696 & 1,000 & 250/750 & 3GPP \\
WCMSA & 312 & 304 & 696 & 1,000 & 250/750 & 3GPP \\
\midrule
\textbf{Total} & 936 & 888 & 2,504 & 3,392 & 848/2,544 & --- \\
\bottomrule
\end{tabular}
\end{table}

\textbf{Data Split Protocol.} The validation set (25\%) is used exclusively for workflow optimization (MCTS search in the reference agent). The test set (75\%) is held out and \emph{never} used for agent tuning, prompt iteration, or hyperparameter selection. This strict separation prevents data leakage.

\textbf{Model Role Separation.} To avoid evaluation bias, we explicitly separate model roles throughout the pipeline:
\begin{itemize}[leftmargin=*]
  \item \textbf{Data Generation / Augmentation:} GPT-4 (2024-04-09)
  \item \textbf{Data Cleaning / Validation:} Multi-model consensus (GPT-4o, Claude-3.5-Sonnet, Qwen-Max, DeepSeek-V3, and others)
  \item \textbf{Data Auditing / Repair:} DeepSeek-R1
  \item \textbf{Evaluation Baselines:} GPT-4o, Qwen-Turbo-Latest (distinct from generation models)
  \item \textbf{Workflow Optimization:} Claude-Opus-4.5 (distinct from data pipeline models)
\end{itemize}

\section{Data Collection Prompts}\label{app:prompts}

We design task-specific prompts for each benchmark component. All system prompts emphasize: (1)~domain accuracy using correct formulas and terminology; (2)~answer verifiability with numerical results that can be validated; and (3)~format compliance following standardized JSONL schemas.

\subsection{WCHW Data Collection Prompt}

\begin{tcolorbox}[breakable, colback=blue!5!white, colframe=blue!75!black, title={System Prompt for WCHW Dataset Collection}]
\footnotesize
\textbf{Role:} You are an expert in wireless communications with extensive knowledge in information theory, digital communications, and signal processing.

\textbf{Task:} Convert textbook problems into structured data samples with detailed Chain-of-Thought (CoT) reasoning steps. If a question contains multiple sub-questions, split it into separate samples.

\textbf{Requirements:}
\begin{itemize}[nosep]
    \item Provide step-by-step derivations with all intermediate calculations
    \item Use correct wireless communication formulas (Shannon capacity, path loss, BER, etc.)
    \item Ensure unit consistency (Hz, dB, W) throughout the solution
    \item Verify numerical answers by substituting back into original equations
\end{itemize}

\textbf{Output Format:} JSONL with fields: \texttt{question}, \texttt{answer}, \texttt{cot}, \texttt{id}. Mathematical expressions in \LaTeX. IDs starting from \texttt{WCHW-test-0001}.
\end{tcolorbox}

\subsection{WCNS Data Collection Prompt}

\begin{tcolorbox}[breakable, colback=blue!5!white, colframe=blue!75!black, title={System Prompt for WCNS Dataset Generation}]
\footnotesize
\textbf{Role:} 5G network engineer specializing in network slicing and resource management.

\textbf{Task:} Generate structured problems requiring agents to: (1) classify service requests into eMBB or URLLC, (2) calculate bandwidth allocation using proportional fairness, and (3) compute throughput via 3GPP CQI-based spectral efficiency.

\textbf{System Configuration:}
\begin{itemize}[nosep]
    \item eMBB Slice: 90~MHz total, 6--20~MHz/user, 1--15 users
    \item URLLC Slice: 30~MHz total, 1--5~MHz/user, 1--10 users
    \item CQI Range: 1--15 (3GPP TS~38.214 Table~5.2.2.1-3)
    \item Carrier: 3.5~GHz (n78 band), SCS: 30~kHz
\end{itemize}

\textbf{Bandwidth Formula:} $B_{\text{new}} = B_{\text{slice}} / (N_{\text{users}} + 1)$; Throughput: $R = B \times \eta_{\text{CQI}}$.
\end{tcolorbox}

\subsection{WCMSA Data Collection Prompt}

\begin{tcolorbox}[breakable, colback=blue!5!white, colframe=blue!75!black, title={System Prompt for WCMSA Dataset Generation}]
\footnotesize
\textbf{Role:} Mobile network optimization expert specializing in mobility prediction and proactive resource allocation.

\textbf{Task:} Generate problems requiring: (1)~Kalman filtering to predict next position, (2)~CQI estimation at predicted position, (3)~slice selection, and (4)~bandwidth allocation satisfying QoS.

\textbf{Kalman Filter Config:}
\begin{itemize}[nosep]
    \item State: $\mathbf{x} = [x, y, v_x, v_y]^\top$
    \item Transition: $\mathbf{F} = \begin{bmatrix} 1 & 0 & \Delta t & 0 \\ 0 & 1 & 0 & \Delta t \\ 0 & 0 & 1 & 0 \\ 0 & 0 & 0 & 1 \end{bmatrix}$, $\Delta t = 1$~s
    \item Process noise std: 0.5~m; Measurement noise std: 0.1~m
    \item History: 5 positions; Predict: 1 step ahead
\end{itemize}

\textbf{Mobility:} Pedestrian (1--2~m/s), Vehicular (5--15~m/s).
\end{tcolorbox}

\subsection{Data Augmentation Prompts}

\begin{tcolorbox}[breakable, colback=gray!10!white, colframe=gray!75!black, title={System Prompt for Data Augmentation (All Tasks)}]
\footnotesize
\textbf{General Principles:}
\begin{itemize}[nosep]
    \item Maintain identical JSON format with all required fields
    \item Cover all specified knowledge points / scenarios
    \item Each sample must have verifiable CoT reasoning
    \item All ground-truth answers must be recomputed numerically (not copied from LLM output)
    \item Avoid duplication; create new problems through parameter variation and type transformation
\end{itemize}

\textbf{Augmentation Strategies:}
\begin{itemize}[nosep]
    \item \textbf{Parameter Substitution:} Systematically replace values while preserving structure
    \item \textbf{Bidirectional Conversion:} given $X \to $ find $Y$ $\Leftrightarrow$ given $Y \to$ find $X$
    \item \textbf{Cross-Topic Integration:} Combine concepts from different knowledge areas
    \item \textbf{Boundary Exploration:} Test extreme cases (very high/low SNR, edge-of-cell positions)
\end{itemize}

\textbf{Quality Control:}
\begin{enumerate}[nosep]
    \item Topical consistency: strictly within wireless communication domain
    \item Logical rigor: every CoT must be independently verifiable
    \item Unit consistency: explicit conversion between kHz/MHz, dB/linear
    \item Difficulty gradient: from basic calculation to comprehensive design
\end{enumerate}
\end{tcolorbox}

\section{Reference Agent (\wbr{}) Implementation Details}\label{app:ref_agent}

\wbr{} is provided as a task-aware reference pipeline rather than a novel method contribution. Because its tool interface, field structure, and workflow were co-designed with the benchmark tasks, its score represents the ceiling achievable by a tightly engineered pipeline with full knowledge of the task structure (see Section~\ref{sec:co_design} for discussion). It is built on the AFlow framework~\cite{zhang2025aflow} with MCTS-based workflow optimization and domain-specific tools, as described in our companion work~\cite{tong2025wirelessagentpp}. This section documents the implementation for reproducibility.

\subsection{Architecture Overview}

The agent architecture consists of two loops:
\begin{itemize}[leftmargin=*]
  \item \textbf{Optimize Loop:} An \emph{optimizer LLM} (Claude-Opus-4.5) maintains a search tree where each node represents a candidate workflow. Through selection (UCB with penalty), expansion (LLM-generated workflow mutations), and backpropagation (three-class experience: success / failure / neutral), the optimizer iteratively discovers better workflows.
  \item \textbf{Execute Loop:} An \emph{executor LLM} (Qwen-Turbo-Latest) runs the candidate workflow on validation data. The evaluation module computes tolerance-aware scores, and results are backpropagated to guide future search.
\end{itemize}

\subsection{Domain-Specific Tools}

\wbr{} integrates the following tools:
\begin{itemize}[leftmargin=*]
  \item \textbf{Python Code Executor:} Executes LLM-generated Python code for numerical verification using \texttt{math}, \texttt{numpy}, and \texttt{scipy.special} (e.g., \texttt{erfc}, Bessel functions, Marcum Q-function).
  \item \textbf{Unit Converter:} Automatically detects and converts between unit families (Table~\ref{tab:unit_conversion}).
  \item \textbf{Formula Verifier:} Retrieves domain formulas from a curated knowledge base of $>$10 formula families (Shannon capacity, BER expressions, Carson's rule, etc.) and verifies structural match.
  \item \textbf{Ray-Tracing Predictor:} Returns exact CQI at given $(x,y)$ coordinates on the digital twin (Appendix~\ref{app:digital_twin}).
  \item \textbf{Kalman Filter Predictor:} Fits a constant-velocity Kalman filter to historical positions and returns predicted next position.
\end{itemize}

\subsection{Discovered Workflow Patterns}

The MCTS optimizer discovers three characteristic workflow patterns, one per tier:

\textbf{WCHW: Reason-then-Verify.}
\begin{enumerate}[nosep]
  \item \texttt{Custom} operator solves the problem using a domain-enriched prompt with $>$10 formula families.
  \item \texttt{ToolAgent} generates Python code to independently recalculate the answer and emits the verified value.
\end{enumerate}

\textbf{WCNS: Tool-then-Reason.}
\begin{enumerate}[nosep]
  \item \texttt{CodeLevelRayTracing} (deterministic, no LLM) extracts coordinates, queries the ray-tracing model, and injects CQI into the problem.
  \item \texttt{Custom} operator classifies intent, allocates bandwidth via proportional fairness, and computes throughput using the injected CQI.
\end{enumerate}

\textbf{WCMSA: Predict-Estimate-then-Reason.}
\begin{enumerate}[nosep]
  \item \texttt{CodeLevelKalmanPredictor} (deterministic) predicts the user's next position from historical trajectory.
  \item \texttt{CodeLevelRayTracing} (deterministic) estimates CQI at the \emph{predicted} (not current) position.
  \item \texttt{Custom} operator performs intent classification, bandwidth allocation, throughput computation, and QoS verification using the predicted position and estimated CQI.
\end{enumerate}

\subsection{MCTS Search Cost}

The entire MCTS optimization is remarkably inexpensive:

\begin{table}[h]
\centering
\caption{MCTS search cost for \wbr{} workflow optimization.}
\label{tab:search_cost_app}
\footnotesize
\begin{tabular}{@{}lccc@{}}
\toprule
& \textbf{WCHW} & \textbf{WCNS} & \textbf{WCMSA} \\
\midrule
Search Rounds & 19 & 11 & 11 \\
Wall-Clock Time (min) & 63 & 13 & 14 \\
Total Search Cost (USD) & 4.95 & 0.99 & 1.05 \\
Per-Problem Inference Cost (USD) & 0.00083 & 0.00056 & 0.00068 \\
\bottomrule
\end{tabular}
\end{table}

\subsection{Ablation Study}

Table~\ref{tab:ablation_app} reports an ablation study on the WCHW validation set, quantifying the contribution of each component.

\begin{table}[h]
\centering
\caption{Ablation study on WCHW validation set. $\Delta$ denotes the change from the full \wbr{} configuration.}
\label{tab:ablation_app}
\footnotesize
\begin{tabular}{@{}lcc@{}}
\toprule
\textbf{Configuration} & \textbf{Accuracy (\%)} & $\boldsymbol{\Delta}$ \\
\midrule
\wbr{} (Full) & \textbf{81.78} & --- \\
\quad w/o Penalized Selection & 79.11 & $-2.67$ \\
\quad w/o Heuristic Critic & 80.34 & $-1.44$ \\
\quad w/o 3-Class Experience & 78.59 & $-3.19$ \\
\quad w/o Domain Tools & 62.44 & $-19.34$ \\
\midrule
AFlow (no enhancements) & 69.92 & $-11.86$ \\
\bottomrule
\end{tabular}
\end{table}

Removing domain tools incurs the largest degradation ($-19.34$~pp), consistent with the observation that tool integration is the dominant capability factor. The three-class experience mechanism prevents the optimizer from chasing noise ($-3.19$~pp without it), and penalized selection avoids wasted exploration ($-2.67$~pp).

\section{Experimental Configuration}\label{app:exp_config}

\subsection{Model Specifications}
Table~\ref{tab:model_specs} summaries the model specifications for reproducibility.
\begin{table}[h]
\centering
\caption{Model specifications for reproducibility.}
\label{tab:model_specs}
\scriptsize
\begin{tabular}{@{}llll@{}}
\toprule
\textbf{Model} & \textbf{API / Version} & \textbf{Role} & \textbf{Access Date} \\
\midrule
GPT-4o & gpt-4o-2024-08-06 & Baseline & 2026-01-15 \\
Qwen-Turbo & qwen-turbo-2024-11-01 & Executor & 2026-01-15 \\
Claude-Opus-4.5 & claude-opus-4-5-20250219 & Optimizer & 2026-02-01 \\
\bottomrule
\end{tabular}
\end{table}

\subsection{Inference Hyperparameters}
Table~\ref{tab:hyperparams} gives the inference hyperparameters for all methods.
\begin{table}[h]
\centering
\caption{Inference hyperparameters for all methods.}
\label{tab:hyperparams}
\scriptsize
\begin{tabular}{@{}lccccl@{}}
\toprule
\textbf{Param.} & \textbf{Direct} & \textbf{CoT-SC} & \textbf{MedPr.} & \textbf{Agent} & \textbf{Notes} \\
\midrule
Temperature & 0.0 & 0.7 & 0.7 & 0.0 & Diversity \\
Top-$p$ & 1.0 & 0.95 & 0.95 & 0.95 & --- \\
Max tokens & 2048 & 2048 & 2048 & 4096 & Agent longer \\
Samples ($k$) & 1 & 5 & 5 & 1 & Per problem \\
Retries & 0 & 0 & 0 & 3 & Tool-call \\
\bottomrule
\end{tabular}
\end{table}

\subsection{Prompt Template}

All methods use a unified base prompt template:
\begin{tcolorbox}[colback=gray!5!white, colframe=gray!75!black, title={Evaluation Prompt Template}]
\footnotesize
\texttt{You are an expert in wireless communications. Solve the following problem step by step. Show your reasoning clearly and provide the final answer in the specified format.}

\texttt{Problem: \{question\}}

\texttt{Required output format: \{format\_spec\}}
\end{tcolorbox}

For CoT-SC and MedPrompt, this base prompt is augmented with chain-of-thought elicitation instructions. For agentic methods (ADAS, AFlow, \wbr{}), the prompt is auto-generated by the optimizer.

\section{Representative Dataset Examples}\label{app:examples}

This section provides detailed examples from each benchmark tier, illustrating problem format, chain-of-thought structure, and evaluation criteria.

\subsection{WCHW Examples: Domain Knowledge Reasoning}

\begin{tcolorbox}[breakable, colback=orange!5!white, colframe=orange!75!black, title={WCHW Example~1: Shannon Capacity (Basic)}]
\footnotesize
\textbf{Question:} A wireless channel has bandwidth $B=50$~MHz and SNR$\,=\,0.1$ (linear). Compute the channel capacity $C$ in Mbps.

\textbf{Reference Answer:} $6.87$~Mbps.

\textbf{Chain-of-Thought:}
\begin{enumerate}[nosep]
  \item Apply Shannon's formula: $C = B\log_2(1+\text{SNR})$.
  \item Substitute: $C = 50\times10^6 \times \log_2(1.1)$.
  \item Compute: $\log_2(1.1) = \ln(1.1)/\ln(2) = 0.09531/0.6931 = 0.1375$.
  \item $C = 50\times10^6 \times 0.1375 = 6.875\times10^6$~bps $= 6.87$~Mbps.
\end{enumerate}

\textbf{Scoring Notes:} Numeric type; tolerance at 1\%: answers in $[6.80, 6.94]$ receive full credit. An answer of ``6.87~kbps'' (unit error) receives score~$=0$.
\end{tcolorbox}

\begin{tcolorbox}[breakable, colback=orange!5!white, colframe=orange!75!black, title={WCHW Example~2: BER for Noncoherent BFSK (Intermediate)}]
\footnotesize
\textbf{Question:} Compute the BER for noncoherent BFSK at $E_b/N_0 = 8$~dB.

\textbf{Reference Answer:} $2.13\times10^{-2}$.

\textbf{Chain-of-Thought:}
\begin{enumerate}[nosep]
  \item Convert $E_b/N_0$ from dB to linear: $\gamma = 10^{8/10} = 10^{0.8} = 6.31$.
  \item Apply noncoherent BFSK BER formula: $P_b = \frac{1}{2}e^{-\gamma/2}$.
  \item Compute exponent: $\gamma/2 = 3.155$.
  \item Compute: $e^{-3.155} = 0.0426$.
  \item $P_b = 0.5 \times 0.0426 = 0.0213 = 2.13\times10^{-2}$.
\end{enumerate}

\textbf{Scoring Notes:} Scientific notation; mantissa tolerance 5\% with exponent match required. Using BPSK formula ($0.5\,\mathrm{erfc}(\sqrt{E_b/N_0})$) instead of noncoherent BFSK is a formula misapplication error.
\end{tcolorbox}

\begin{tcolorbox}[breakable, colback=orange!5!white, colframe=orange!75!black, title={WCHW Example~3: FM Bandwidth (Carson's Rule)}]
\footnotesize
\textbf{Question:} An FM signal has maximum frequency deviation $\Delta f = 75$~kHz and maximum modulating frequency $f_m = 15$~kHz. Compute the transmission bandwidth using Carson's rule.

\textbf{Reference Answer:} $180$~kHz.

\textbf{Chain-of-Thought:}
\begin{enumerate}[nosep]
  \item Apply Carson's rule: $\text{BW} = 2(\Delta f + f_m)$.
  \item Substitute: $\text{BW} = 2(75 + 15) = 2 \times 90 = 180$~kHz.
\end{enumerate}

\textbf{Scoring Notes:} Common error: using $\text{BW}=2\Delta f$ (narrowband approximation) gives 150~kHz---17\% error, scored 0.
\end{tcolorbox}

\subsection{WCNS Examples: Intent-to-Allocation Decision}

\begin{tcolorbox}[breakable, colback=blue!5!white, colframe=blue!60!black, title={WCNS Example~1: eMBB Web Browsing}]
\footnotesize
\textbf{Network State:}
\begin{itemize}[nosep,leftmargin=*]
  \item eMBB Slice: 12 active users, 90~MHz total capacity
  \item URLLC Slice: 3 active users, 30~MHz total capacity
\end{itemize}

\textbf{New User:} Position $(45.2, 78.1)$, Region: North, Service Request: ``I want to browse websites and check email''

\textbf{Reference Output:}
\begin{itemize}[nosep,leftmargin=*]
  \item Slice Type: eMBB
  \item CQI: 8 (from ray-tracing tool)
  \item Bandwidth: $90/(12+1) = 6.92$~MHz
  \item Throughput: $6.92 \times 1.91 = 13.2$~Mbps
\end{itemize}

\textbf{Chain-of-Thought:}
\begin{enumerate}[nosep]
  \item \textbf{Intent Classification:} Web browsing and email are throughput-oriented, latency-tolerant services $\to$ eMBB.
  \item \textbf{CQI Acquisition:} Call \texttt{ray\_tracing(x=45.2, y=78.1, region=``North'')} $\to$ CQI~$=8$.
  \item \textbf{Bandwidth Allocation:} Proportional fairness with $N+1$ users: $B = 90/(12+1) = 6.92$~MHz.
  \item \textbf{Throughput:} CQI~$=8$ $\to$ $\eta = 1.91$~bps/Hz (3GPP Table~\ref{tab:cqi}). $R = 6.92 \times 1.91 = 13.2$~Mbps.
\end{enumerate}

\textbf{Scoring Notes:} The CQI field is the most critical---without calling the ray-tracing tool, agents must guess CQI, which is scored near zero.
\end{tcolorbox}

\begin{tcolorbox}[breakable, colback=blue!5!white, colframe=blue!60!black, title={WCNS Example~2: URLLC Industrial Control}]
\footnotesize
\textbf{Network State:}
\begin{itemize}[nosep,leftmargin=*]
  \item eMBB Slice: 8 active users, 90~MHz total capacity
  \item URLLC Slice: 5 active users, 30~MHz total capacity
\end{itemize}

\textbf{New User:} Position $(23.7, 51.3)$, Region: Center, Service Request: ``We need to control factory robots with real-time commands''

\textbf{Reference Output:}
\begin{itemize}[nosep,leftmargin=*]
  \item Slice Type: URLLC
  \item CQI: 12 (from ray-tracing tool)
  \item Bandwidth: $30/(5+1) = 5.0$~MHz
  \item Throughput: $5.0 \times 3.90 = 19.5$~Mbps
\end{itemize}

\textbf{Chain-of-Thought:}
\begin{enumerate}[nosep]
  \item \textbf{Intent Classification:} Factory robot control requires ultra-low latency $\to$ URLLC.
  \item \textbf{CQI Acquisition:} Call ray-tracing tool at $(23.7, 51.3)$ in Center region $\to$ CQI~$=12$.
  \item \textbf{Bandwidth:} $B = 30/(5+1) = 5.0$~MHz (clamped to $[1,5]$ range).
  \item \textbf{Throughput:} CQI~$=12$ $\to$ $\eta = 3.90$~bps/Hz. $R = 5.0 \times 3.90 = 19.5$~Mbps.
\end{enumerate}
\end{tcolorbox}

\subsection{WCMSA Examples: Proactive Mobile Service Assurance}

\begin{tcolorbox}[breakable, colback=green!5!white, colframe=green!60!black, title={WCMSA Example: Cloud Gaming Under Mobility}]
\footnotesize
\textbf{Input:}
\begin{itemize}[nosep,leftmargin=*]
  \item Historical positions: $(79.3, 46.0) \to (80.1, 45.4) \to (81.2, 44.7) \to (82.1, 44.1)$
  \item Base Station: Location $(0, 0)$, Tx power: 46~dBm
  \item Network State: eMBB: 2 active users (90~MHz); URLLC: 1 active user (30~MHz)
  \item Service Request: ``Cloud gaming with high graphics,'' Min rate: 35~Mbps
\end{itemize}

\textbf{Reference Output (6 fields):}
\begin{itemize}[nosep,leftmargin=*]
  \item Predicted Position: $(83.0, 43.5)$
  \item CQI: 15 (at predicted position, from ray-tracing)
  \item Slice Type: eMBB
  \item Bandwidth: $\min(90/3, 20) = 20$~MHz
  \item Throughput: $20 \times 5.55 = 111.0$~Mbps
  \item QoS Satisfied: Yes ($111.0 > 35$~Mbps)
\end{itemize}

\textbf{Chain-of-Thought:}
\begin{enumerate}[nosep]
  \item \textbf{Position Prediction:} Apply Kalman filter. Velocity estimate: $v_x \approx 82.1 - 81.2 = 0.9$~m/s, $v_y \approx 44.1 - 44.7 = -0.6$~m/s. Predicted: $\hat{x} = 82.1 + 0.9 = 83.0$, $\hat{y} = 44.1 - 0.6 = 43.5$.
  \item \textbf{CQI Estimation:} Call \texttt{ray\_tracing(x=83.0, y=43.5, region=``South'')} $\to$ CQI~$=15$.
  \item \textbf{Intent Classification:} Cloud gaming is high-throughput $\to$ eMBB.
  \item \textbf{Bandwidth:} $B = 90/(2+1) = 30$~MHz, clamped to max 20~MHz.
  \item \textbf{Throughput:} CQI~$=15$ $\to$ $\eta = 5.55$~bps/Hz. $R = 20 \times 5.55 = 111.0$~Mbps.
  \item \textbf{QoS Check:} $111.0 \geq 35$~Mbps $\to$ QoS Satisfied: Yes.
\end{enumerate}

\textbf{Scoring Notes:} Position error propagates through CQI estimation. A 5~m prediction error can shift CQI by 2--3 levels, cascading to bandwidth, throughput, and QoS fields.
\end{tcolorbox}

\section{Copyright Compliance and Licensing}\label{app:license}

\textbf{Data Provenance.} All problems in WirelessBench are \emph{synthetically generated} based on standard engineering concepts and publicly available specifications. No copyrighted textbook content is directly reproduced. WCHW problems are \emph{inspired by} canonical wireless communication concepts (Shannon capacity, modulation theory, etc.); only underlying mathematical relationships are used, with original problem text newly composed. WCNS and WCMSA problems are programmatically generated from 3GPP TS~38.214 specifications, which are publicly available technical standards.

\textbf{Licensing.} WirelessBench will be released under CC BY-NC~4.0, permitting academic use with attribution. The evaluator code and scoring scripts are open-sourced under MIT license.

\textbf{Ethical Statement.} The benchmark contains no personally identifiable information. All network scenarios are synthetic and do not represent real deployments. The benchmark is designed for evaluation purposes and should not be used as a direct replacement for certified network-control pipelines.

\section{Reproducibility Checklist}\label{app:repro}

To enable full reproduction of all results reported in this paper, we release the following artifacts at \url{https://github.com/jwentong/WirelessBench}:

\begin{enumerate}[leftmargin=*]
  \item \textbf{Benchmark Data:} All 3,392 samples in JSONL format with fixed train/val/test splits, including question text, reference answer, chain-of-thought solution, metadata (task type, knowledge category, difficulty level).
  \item \textbf{Scoring Scripts:} Complete evaluator implementation including answer-type classification, unit normalization, tolerance-aware scoring, and catastrophic-error detection.
  \item \textbf{Prompt Templates:} All evaluation prompts for direct prompting, CoT-SC, and MedPrompt baselines.
  \item \textbf{Ray-Tracing Data:} Pre-computed propagation maps for North/Center/South regions of the HKUST radio map.
  \item \textbf{Kalman Filter Implementation:} Deterministic position predictor for WCMSA trajectory prediction.
  \item \textbf{Experiment Logs:} Raw model outputs, parsed answers, and per-sample scores for all results reported in this paper, with model version identifiers and timestamps.
  \item \textbf{Statistical Analysis Scripts:} Bootstrap confidence interval computation, significance testing, and effect size calculation.
\end{enumerate}

\textbf{Minimum Reproduction Requirements:}
\begin{itemize}[leftmargin=*]
  \item Fixed model version (recorded in Table~\ref{tab:model_specs})
  \item Fixed prompt template version (commit hash)
  \item Fixed decoding parameters (Table~\ref{tab:hyperparams})
  \item Deterministic evaluator version (release tag)
\end{itemize}

\end{document}